\documentclass[aip,
jcp,
 amsmath,amssymb,
 reprint,
 longbibliography]{revtex4-1}
\usepackage[utf8]{inputenc}
\usepackage[version=4]{mhchem}
\usepackage{xcolor}
\usepackage{tabularx}
\usepackage{booktabs}
\usepackage{graphicx}
\usepackage{tablefootnote}
\usepackage{braket}
\usepackage{siunitx}
\DeclareSIUnit{\angstrom}{\mbox{\normalfont\AA}}

\draft 

\begin{document}
\title{The binding of atomic hydrogen on graphene from density functional theory and diffusion Monte Carlo calculations}

\author{Amanda Dumi}
\affiliation{Department of Chemistry, University of Pittsburgh, Pittsburgh, Pennsylvania 15260,
USA}
\author{Shiv Upadhyay}
\affiliation{Department of Chemistry, University of Pittsburgh, Pittsburgh, Pennsylvania 15260,
USA}
\author{Leonardo Bernasconi}
\affiliation{Department of Chemistry, University of Pittsburgh, Pittsburgh, Pennsylvania 15260,
USA}
\affiliation{Center for Research Computing,
University of Pittsburgh, Pittsburgh, Pennsylvania 15260, USA}
\author{Hyeondeok Shin}
\affiliation{Computational Science Division, Argonne National Laboratory, Argonne, Illinois 60439, USA}
\author{Anouar Benali}
\affiliation{Computational Science Division, Argonne National Laboratory, Argonne, Illinois 60439, USA}
\author{Kenneth D. Jordan}
\affiliation{Department of Chemistry, University of Pittsburgh, Pittsburgh, Pennsylvania 15260,
USA}
\affiliation{Department of Chemical and Petroleum Engineering, University of Pittsburgh, Pittsburgh, Pennsylvania 15260,
USA}
\date{\today}








\begin{abstract}
In this work density functional theory (DFT) and diffusion Monte Carlo (DMC) methods are used to calculate the binding energy of a H atom chemisorbed on the graphene surface.
The Perdew-Burke-Ernzerhof (PBE) value of the binding energy is about 20\% larger in magnitude than the diffusion Monte Carlo result.
The inclusion of exact exchange through the use of the Heyd–Scuseria–Ernzerhof (HSE) functional brings the DFT value of the binding energy closer in line with the DMC result.
It is also found that there are significant differences in the charge distributions determined using PBE and DMC approaches.

\end{abstract}
\maketitle
\section{Introduction}

The unique electronic, optical, and transport properties of graphene make it an important system for a wide range of applications, many of which involve or are impacted by the adsorption of atoms or molecules.

To bring these applications to fruition, a deeper understanding of the interaction of atoms and molecules with graphene is required, and, not surprisingly, this has been the subject of several experimental and theoretical studies. \cite{doi:10.1021/acs.jpca.0c06595,10.1016/j.carbon.2021.02.056,10.1063/1.3569134,shin_diffusion_2019,10.1103/PhysRevB.84.033402, D1CP02473F,doi:10.1126/science.1158877,10.1038/nature04233,Alekseeva2020,10.1016/j.rser.2015.05.011,reactions2030014,doi:10.1063/1.2963976,doi:10.1021/nl801417w}

The adsorption of H atoms on graphene has been the subject of multiple studies.\cite{doi:10.1016/j.cplett.2010.10.1002/wcms.136010.027,10.1063/1.3569134,10.1016/j.rser.2015.05.011,reactions2030014,doi:10.1063/1.2963976,doi:10.1021/nl801417w}
It is known that there is both a weakly absorbed state in which barriers for diffusion are small and a much more strongly bound chemisorbed state. \cite{SHA2002318,JELOAICA1999157}
Chemisorbed H atoms open up the band gap and allow for tuning of electronic properties. \cite{10.1007/s40089-017-0203-5}
It has been demonstrated that even a single chemisorbed hydrogen atom causes an extended magnetic moment in the graphene sheet.\cite{gonzalez-herrero_atomic-scale_2016,GonzlezHerrero2019}
On the other hand, there is evidence that given the ready diffusion of H in the physisorbed state, the H atoms tend to pair up on the surface leading to non-magnetic species.\cite{10.1088/2053-1583/ab03a0}
Finally, interest in the hydrogen/graphene system has  also been motivated by the potential use of graphene and graphitic surfaces for hydrogen storage.\cite{Alekseeva2020}

The majority of computational studies of adsorption of atoms and molecules on graphene have employed density functional theory (DFT), primarily due to its favorable scaling with system size, allowing for the treatment of larger periodic structures.
However, a reliable theoretical description of interactions at the graphene surface has proven to be challenging for DFT.\cite{doi:10.1021/acs.jpca.0c06595,10.1016/j.carbon.2021.02.056,10.1063/1.3569134,doi:10.1063/1.4977994}
In recent years considerable progress has been made in extending correlated wave function methods to periodic systems. \cite{doi:10.1063/1.5091445,doi:10.1063/5.0049890,10.1038/nature11770,doi:10.1063/5.0036363, doi:10.1063/5.0021036,doi:10.1063/1.4976937}
Among these methods, the diffusion Monte Carlo (DMC)\cite{foulkes01} method, which  is a real-space stochastic approach to solving the many-body Schr{\"o}dinger equation is particularly attractive given its low scaling with the number of electrons and high parallelizability.
DMC also has the advantages of being systematically improvable and being much less sensitive to the basis set employed than methods that work in the space of Slater determinants.
DMC has been used to describe the adsorption of various species on graphene including \ce{O2}\cite{shin_diffusion_2019}, a water molecule\cite{10.1103/PhysRevB.84.033402,doi:10.1021/acs.jpclett.8b03679}, and a platinum atom.\cite{D1CP02473F}
In a study of a physisorbed H atom on graphene, Ma et al.~found that different DFT functionals gave binding energies ranging from 5 to 97 meV, while DMC calculations gave a value of only 5 $\pm$ 5 meV.\cite{10.1063/1.3569134}
Various DFT calculations utilizing the Perdew-Burke-Ernzerhof (PBE)\cite{10.1103/PhysRevLett.77.3865} and Perdew-Wang (PW91)\cite{PW91} functionals predict the chemisorbed H atom species to be bound by 480 to 1,440 meV.\cite{10.1016/j.carbon.2006.09.027, doi:10.1063/1.3187941, 10.1103/PhysRevLett.93.187202, 10.1103/PhysRevB.78.041402, doi:10.1063/1.3072333, 10.1088/0957-4484/19/15/155708,10.1088/1742-6596/100/5/052087, 10.1016/j.jmmm.2009.11.014,PhysRevB.77.035427}
However, this large spread is primarily a result of some calculations employing small supercells resulting in an unphysical description of the low-coverage situation, too small a $k$-point grid, or small atom-localized basis sets that do not adequately describe the binding and introduce large basis set superposition error (BSSE).
In the present work, we use the DMC method to calculate the binding energy of H to graphene in the chemisorbed state.

\section{Methods}
All calculations reported in this study used a 5x5x1 supercell of graphene,
as it was large enough to make inconsequential the interaction between periodic images of the adsorbed hydrogen atom and to assure that there are essentially unperturbed C atoms between the buckled regions in adjacent images in the $x$ and $y$ directions.
The geometries of graphene, both pristine and with a hydrogen adsorbate, were provided by Kim et al.,\footnote{ Under review: M. A. Kim, D. Sorescu, S. Amemiya, K. D. Jordan, and H. Liu, ``Real Time Modulation of Hydrogen Evolution Activity of Graphene Electrodes Using Mechanical Strain,'' \textit{ACS Appl. Mater. Interfaces}.} and were obtained using the PBE+D3 DFT method.\cite{doi:10.1063/1.3382344,10.1103/PhysRevLett.77.3865}
For all systems, a vacuum spacing of \SI{16}{\angstrom} was used.

\subsection{Density functional theory calculations}
The single particle orbitals used in the trial wave functions for variational Monte Carlo (VMC) and DMC calculations were calculated using the PBE functional with the core-correlated electron core potential (ccECP)\cite{ccecp_1,ccecp_2} pseudopotentials and a plane wave basis with an energy cutoff of 3,400 eV.
Monkhorst-Pack $k$-point grid meshes\cite{PhysRevB.13.5188} were employed with a 13.6 meV Marzari-Vanderbilt-DeVita-Payne cold smearing of the occupations.\cite{PhysRevLett.82.3296}
The PBE results were converged at a 6x6x1 $k$-point grid.

In addition to the PBE calculations used to generate the trial wave functions for DMC, DFT calculations were carried out with the PBE0\cite{doi:10.1063/1.478522} and Heyd–Scuseria–Ernzerhof (HSE) functionals\cite{doi:10.1063/1.2404663} to determine if inclusion of exact exchange proves important for the adsorption energy.
Due to the inclusion of exact exchange, these calculations would be computationally demanding in a plane wave basis, particularly with the high energetic cutoff required by the ccECP pseudopotential.
For this reason, they were carried out all-electron with the POB-TZVP Gaussian type orbital (GTO) basis set.\cite{Peintinger2012-ff}
Due to the use of GTOs, these calculations suffer from basis set superposition error (BSSE), for which we use Grimme's geometry-dependent counterpoise correction.\cite{KruseGrimme2012,Brandenburg2013}
For the PBE0 and HSE, a 12x12x1 $k$-point grid was used to assure well converged energies.

\subsection{Quantum Monte Carlo calculations}

DMC is a projector quantum Monte Carlo (QMC) method, solving the Schr\"{o}dinger equation in imaginary time $\tau=it$; any initial state $\ket{\psi}$, that is not orthogonal to the true ground state $\ket{\phi_0}$ , will evolve to the ground state in the long time limit.
When dealing with Fermionic particles, the DMC method requires the use of the fixed-node approximation\cite{Anderson1980} to maintain the antisymmetric property of the wave function.
For efficient sampling and to reduce statistical fluctuations, we use a Slater-Jastrow trial wave function fixing the nodes through a Slater determinant comprised of single-particle orbitals, which, in this work, are expanded in a B-spline basis.
The Jastrow factor is a function that reduces the variance by explicitly describing dynamic correlation.
The Jastrow factor contains terms for one-body (electron-ion), two-body (electron-electron) and three-body  (electron-electron-ion) interactions.
In this study, 10 parameters were employed per spin-channel and the cutoff was fixed to the Wigner-Seitz radius of the simulation cell.
The parameters in the Jastrow function were optimized with the linear method\cite{10.1103/PhysRevLett.98.110201} using VMC.
To reduce the cost of the DMC calculations as well as to reduce the fluctuations near the ionic core regions, ccECP pseudopotentials were used to replace the core electrons.\cite{ccecp_1,ccecp_2}
The ccECP pseudopotentials were designed to be used with high-accuracy many-body methods such as DMC.
\begin{figure}
    \centering
    \includegraphics[width=\columnwidth,keepaspectratio]{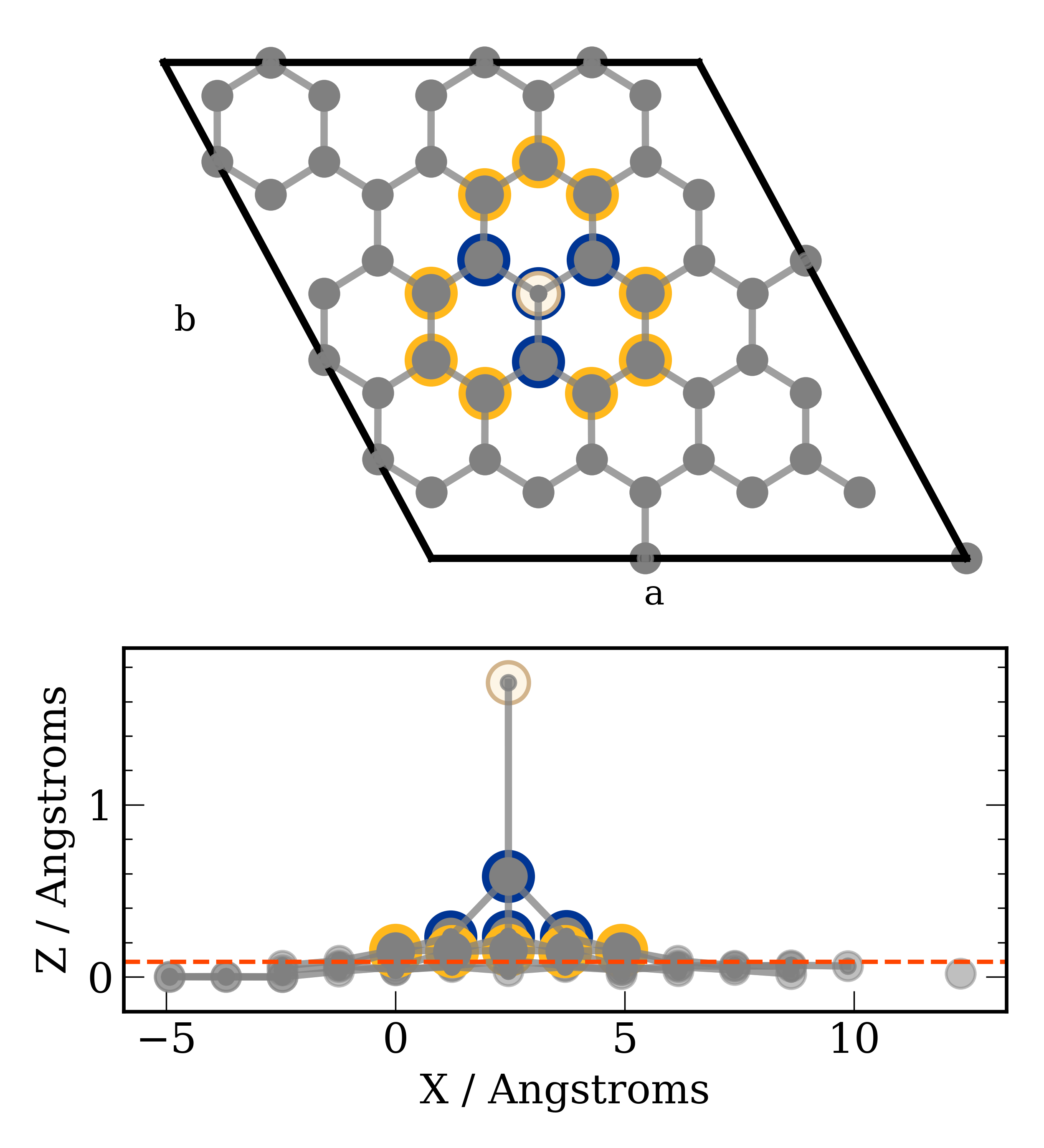}
    \caption{Perpendicular view of the simulation cell (top) and a parallel view obtained by projection onto the $xz$-plane (bottom). The carbon atoms are colored gray and the hydrogen atom is denoted as white. For the parallel view, the orange line represents the mean carbon $z$ position. Blue outlined atoms are greater than one standard deviation away from the mean carbon $z$ position, whereas yellow atoms are between 0.5-1.0 $\sigma$.}
    \label{fig:cell}
\end{figure}
The non-local effects due to the ccECP pseudopotentials were addressed using the determinant-localization approximation along with the t-moves method (DLTM).\cite{zen_new_2019, 10.1063/1.3380831}
Finite size effects were addressed using twist averaging, and symmetry unique twist angles were used.\cite{PhysRevE.64.016702}

The DMC calculations were performed using the branching scheme proposed by Zen
et al. (ZSGMA)\cite{10.1103/PhysRevB.93.241118} with a population control target of 8,192 walkers and a time step of 0.005 a.u., which represented a balance between computational cost and finite timestep error in previous work.\cite{10.1103/PhysRevB.100.075430}

We define the binding energy as,
\begin{equation}
E_b=E_{dgr+H}-(E_{gr}+E_{H})
\label{eq:binding}
\end{equation}
where $E_{dgr+H}$ is the energy of the distorted graphene sheet with a chemisorbed atomic hydrogen, $E_H$ is the energy of a hydrogen atom, and $E_{gr}$ is the energy of a pristine graphene sheet.
In the chemisorbed state, the hydrogen atom bonds directly over a carbon atom, causing this carbon to be pulled out of the sheet towards the hydrogen.\cite{doi:10.1063/1.4896611,doi:10.1126/science.aaw6378}
The adjacent carbons are also pulled in the direction of the hydrogen leading to a distorted graphene sheet.

The plane wave DFT calculations were carried out with the QUANTUM ESPRESSO version 6.3 code.\cite{doi:10.1063/5.0005082,QE-2009,QE-2017}
The Gaussian basis DFT calculations were carried out with CRYSTAL17,\cite{10.1002/wcms.1360,doi:10.1063/5.0004892} save for the HSE calculation of the lone hydrogen atom which was carried out in NWChem version 6.8\cite{NWCHEM} using the same basis as the calculations in CRYSTAL17.
The QMC calculations were carried out using the QMCPACK code, with the workflow between QUANTUM ESPRESSO and QMCPACK managed by Nexus.\cite{QMCPACK_1,QMCPACK_2,10.1016/j.cpc.2015.08.012}
Figures~\ref{fig:cell} and \ref{fig:densdiff_dmcminusdft} were rendered with matplotlib\cite{matplotlib} and the density plots were generated using VESTA.\cite{10.1107/S0021889811038970}

\section{Results \& Discussion}
\subsection{Binding energy}
 \begin{table}[ht]
     \caption{Binding energy (meV) of a hydrogen atom chemisorbed on graphene calculated with various DFT functionals and with DMC.}
    \centering
    \begin{tabular}{lr}
    Method & Binding energy\\\hline \\\\[-2em]%
    \multicolumn{2}{c}{\bfseries This Work} \\
    PBE \footnote{\label{pw}Calculation was done in the plane wave basis} & -820\\ 
    PBE\footnote{\label{gbs}Calculation was done in the Gaussian basis set with correction for BSSE. Values in parentheses include a correction for the Gaussian basis incompleteness error. See text for details.} & -871\\
    PBE0\textsuperscript{\ref{gbs}}& -851 (-800)\\
    HSE\textsuperscript{\ref{gbs}}& -794 (-743)\\
    DMC  & -691 $\pm$ 19 \\ \hline \\\\[-2em]
    \multicolumn{2}{c}{\bfseries Previous Work}\\
    PW91  &  -810 to -830\cite{doi:10.1063/1.3187941}, -870\cite{10.1103/PhysRevLett.93.187202}\\
    PBE  & -790\cite{10.1103/PhysRevB.78.041402}, -840\cite{doi:10.1063/1.3072333}, -980\cite{10.1088/0957-4484/19/15/155708}\\
     \hline
    \end{tabular}
\label{tab:summary_energetics}
 \end{table}
Table~\ref{tab:summary_energetics} contains a summary of the binding energies of a hydrogen atom chemisorbed on graphene from this work and selected values from previous publications using the PW91 and PBE functionals.
These literature values range from -790 to -z980 meV.
However, this wide spread is largely caused by (1) the use in some studies of small supercells for which there are sizable interactions between the CH groups in adjacent cells, and (2) the use in some studies of small atom-centered basis sets without corrections for BSSE.
Our calculations with the PBE functional in conjunction with a plane wave basis set give a binding energy of -821 meV.
This should be contrasted with our -691 $\pm$ 19 meV DMC result.
There are several possible sources for the difference between the PBE and DMC values of the binding energy.
These include errors in the DFT calculations due to self interaction and planar graphene having more multiconfigurational character than H/graphene, with this being better described with DMC than with PBE.
The PBE binding energy is 51 meV lower in magnitude in the plane wave than in the GTO basis set when the same $k$-point grid is used, and this value is used as a correction for the basis set incompleteness error for the results with other functionals in Table~\ref{tab:summary_energetics}.
The calculations in the GTO basis set give a slightly smaller in magnitude binding energy with PBE0 than with PBE.
However, with HSE, we obtain a binding energy 77 meV smaller in magnitude than the PBE result.
Applying the correction for the basis set incompleteness error, we obtain -800 meV for the PBE0 binding energy and -743 meV for the HSE binding energy, with the latter being in reasonable agreement with the DMC result of -691 meV.

\subsection{Binding density}
\begin{figure*}
    \centering
    \includegraphics[width=\textwidth,keepaspectratio]{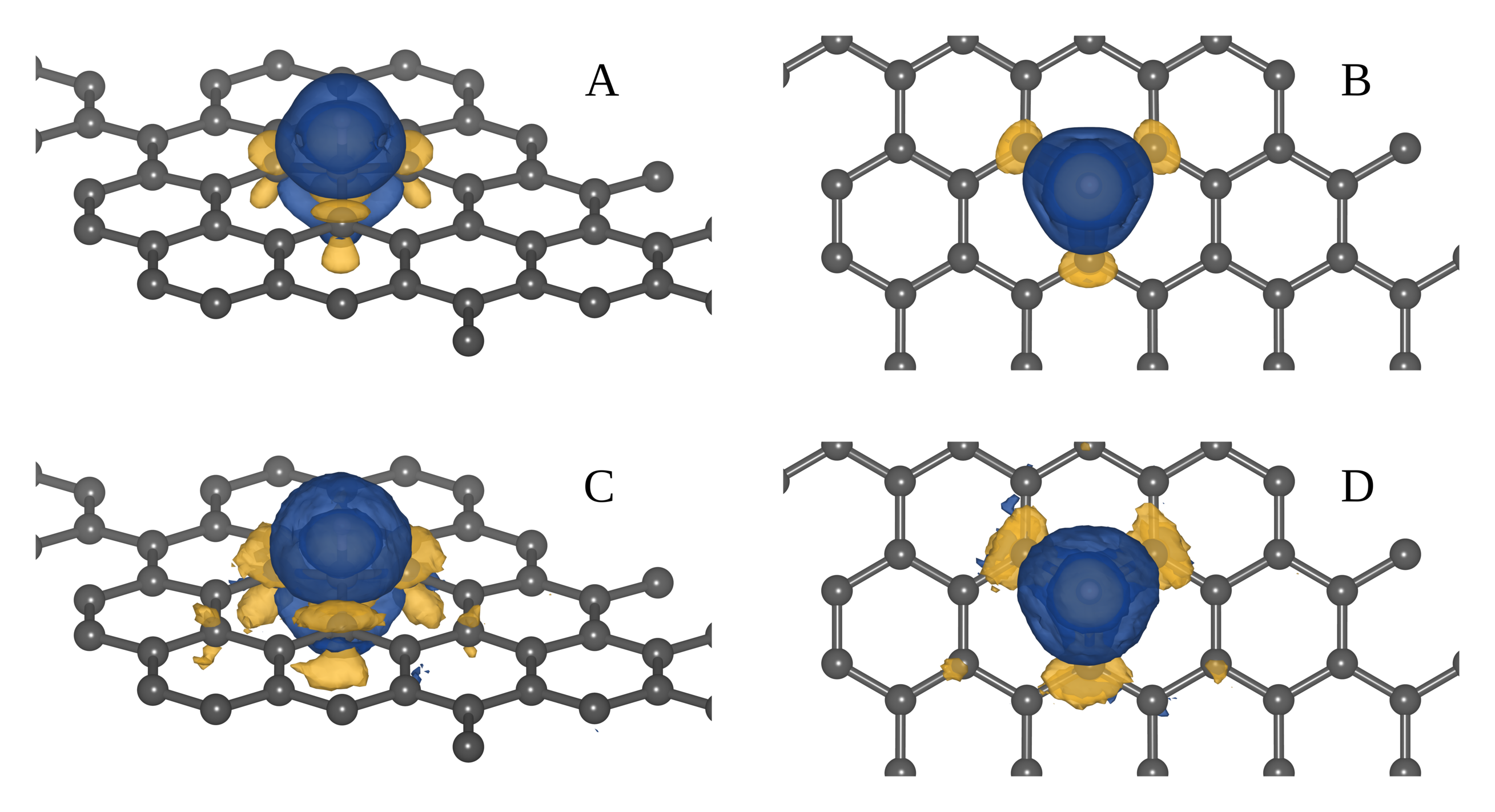}
    \caption{Change of the electron density due to the adsorption of the H atom to the distorted graphene sheet (Eq.~\ref{eq:bindingdensity}). $\rho_b$ from PBE calculations is shown from an oblique angle (A) and aligned along the $c$ axis (B). $\rho_b$ from DMC calculations (C) and (D) is shown from the same perspectives. Gold and blue represent a gain and loss of electron density, respectively. Note that there is a region of increased charge density at the C-H bond that is enveloped by a region of loss in the charge density. The binding density was visualized using an isovalue of 2.8$\times10^{-5}$ for DMC and 3.9$\times10^{-5}$ for PBE, in both cases capturing 95\% of the differential charge density.}
    \label{fig:densdiff}
\end{figure*}
It is instructive to examine the change in the electron density associated with the binding of the H atom to the distorted graphene as determined from the PBE and DMC calculations.
The density change is given by
\begin{equation}
\rho_b=\rho_{dgr+H}-(\rho_{dgr}+\rho_{H}),
\label{eq:bindingdensity}
\end{equation}
where $\rho_{H}$ is the charge density of the hydrogen atom, and $\rho_{dgr+H}$ and $\rho_{dgr}$ are the charge densities of the distorted graphene sheet with and without hydrogen, respectively.

The $\rho_b$ density differences for both DMC and PBE are shown in Figure \ref{fig:densdiff}.
The dark blue and gold regions represent a loss and gain of electron density, respectively.
As expected, there is a shift in electron density from the carbon atom participating in the carbon-hydrogen bond as well as to the three adjacent carbon atoms.
These qualitative changes in the density are consistent with previous theoretical and experimental studies.\cite{doi:10.1063/1.4896611,doi:10.1126/science.aaw6378}
The rehybridization from $sp^2$ to $sp^3$ of the carbon participating in the CH bond and the weakening of the $\pi$ bonds due to the distortion of the graphene lead to the electron density shift.
The change in the charge distribution is similar for PBE and DMC, with the most noticeable difference being a greater increase of density at remote C atoms in the DMC than in the PBE calculations.
Despite the qualitative agreement between the binding density resulting from DMC and PBE, we find that the density differences for the individual systems can offer further insight to the performance of the two methods, as will be seen in the next subsection.

\subsection{Charge density differences between DMC and PBE}
\begin{figure}
    \centering
    \includegraphics[width=\columnwidth,keepaspectratio]{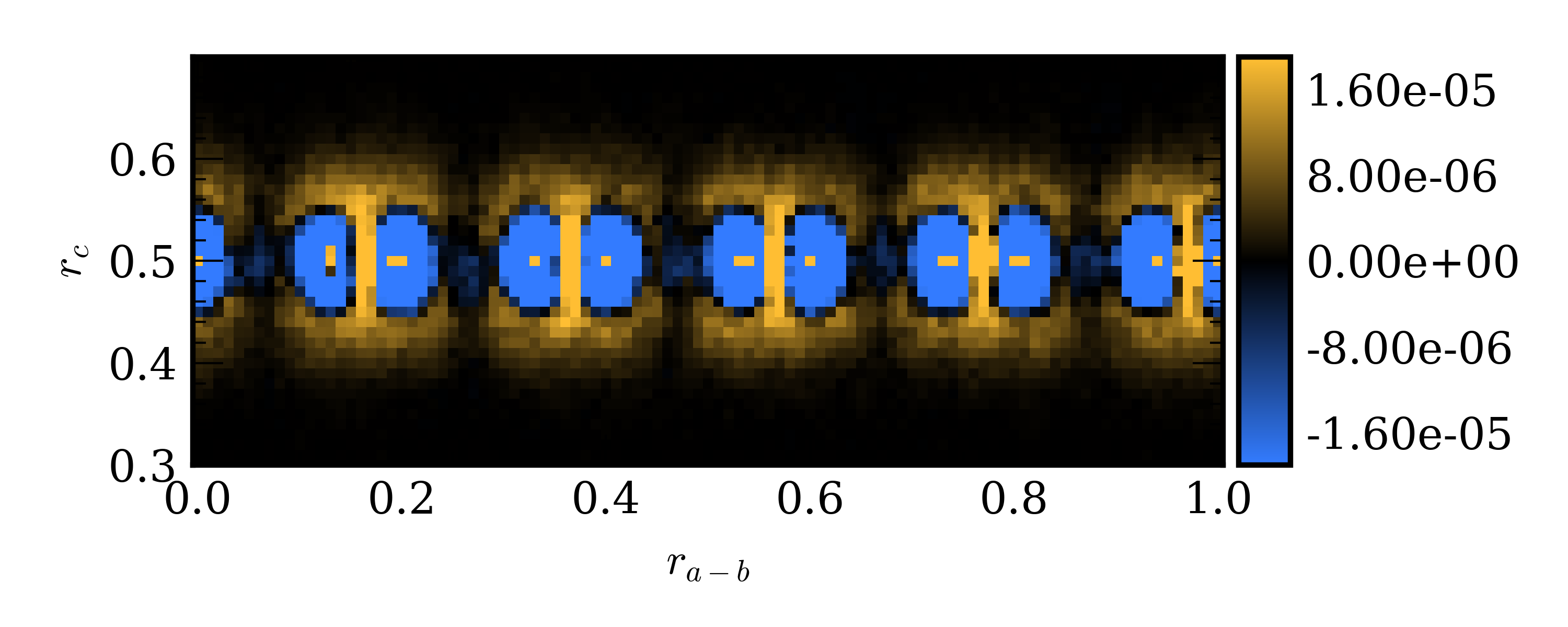}
    \includegraphics[width=\columnwidth,keepaspectratio]{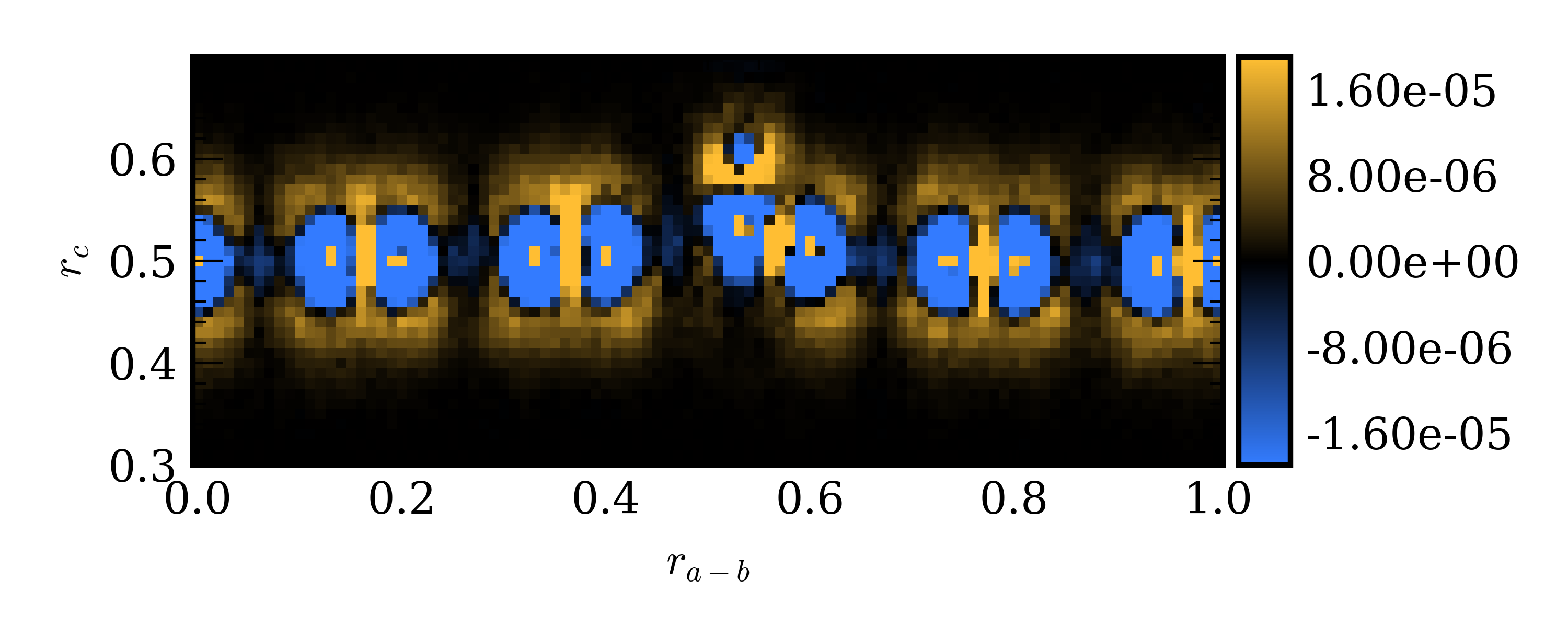}
    \caption{Visualization of the difference of PBE and DMC densities sliced along the 110 lattice plane of the unit cell for the graphene sheet, $\Delta\rho_{gr}$, (top) and H adsorbed onto graphene, $\Delta\rho_{dgr+H}$, (bottom). The abscissa represents traversing the 110 plane in fractional coordinates, while the ordinate represents traversing the $c$ axis in fractional coordinates. Blue regions represent places where the PBE density is larger, while the gold color represents regions where the DMC density is larger. }
    \label{fig:densdiff_dmcminusdft}
\end{figure}

In this section, the difference between the DMC and PBE charge densities for distorted graphene with the adsorbed hydrogen atom as well as for pristine planar graphene without the adsorbed hydrogen atom are considered.
The charge density difference for each system is calculated according to
\begin{equation}
    \Delta\rho_{system} =\rho_{system}^{DMC}- \rho_{system}^{PBE},
    \label{eq:densdiff}
\end{equation}
where $\rho_{system}^{DMC}$ is the DMC charge density of a given system (either distorted graphene with the adsorbed hydrogen or pristine graphene) and $\rho_{system}^{PBE}$ is the corresponding PBE charge density.
$\Delta\rho_{gr}$ and $\Delta\rho_{dgr+H}$ are reported in Figure~\ref{fig:densdiff_dmcminusdft} along the 110 slice through the unit cell, which captures the carbon-hydrogen bond.
From the top-down perspective in Figure~\ref{fig:cell}, the 110 lattice plane bisects the cell diagonally through the longer of the two diagonals.
In Figure~\ref{fig:densdiff_dmcminusdft}, blue represents areas where the PBE density is larger, while gold areas represent areas where the DMC density is larger.
The DMC density, in comparison with the PBE density, has greater weight in the bonding region between atoms.
This is the case for both the planar graphene without hydrogen and the system with hydrogen chemisorbed to graphene.

Even though there are significant differences between the PBE and DMC densities for both systems, the difference is similar in the two systems, consistent with it not introducing a large error in the PBE value of the binding energy.

\section{Conclusions}
Calculations of the binding energy of a hydrogen atom on a graphene sheet were carried out using various DFT methods and with DMC.
The DMC calculations provide a benchmark value of the binding energy.

Our best estimate of the binding energy from DMC calculations is -691 $\pm$ 19 meV.
The PBE result obtained with a plane-wave basis set gives a binding energy about 20\% larger in magnitude than the DMC result.
The global hybrid functional, PBE0, gives a binding energy close to that of PBE.
In comparison, HSE, a range-separated hybrid functional, gives a smaller binding energy of -743 meV, after a correction applied for the basis set incompleteness error, and is much closer to the value of the DMC.
Interestingly, there are significant differences in the DMC and PBE charge densities of both graphene and H/graphene.
Most studies of surface adsorption comparing DMC and DFT results have focused on structures and binding energies.
Our work suggests that in terms of understanding performance of DFT methods it is also important to compare the charge densities.
The importance of the characterization of electron densities by density functionals was the topic of a recent paper by Medevev et al.\cite{doi:10.1126/science.aah5975}

\section*{Data Availability Statement}

The supplementary material document includes the total energies and error bars for the quantum Monte Carlo calculations, the total energies for the DFT calculations, and details of the convergence of the DFT total energies with respect to the $k$-point grid and kinetic energy cutoff of the plane wave basis. The data that support the findings of this study are openly available on the Materials Database Facility at \url{https://acdc.alcf.anl.gov/mdf/detail/dumi_dmc_hgraphene_v1.3}, with the following DOI: 10.18126/s1wc-tya.

\section*{Acknowledgements}
We thank Dr. Dan Sorescu for helpful discussion and for sharing the coordinates of his calculations.
A.B. and H.S were supported by the U.S. Department of Energy, Office of Science, Basic Energy Sciences, Materials Sciences and Engineering Division, as part of the Computational Materials Sciences Program and Center for Predictive Simulation of Functional Materials.
An award of computer time was provided by the Innovative and Novel Computational Impact on Theory and Experiment (INCITE) program.
The DMC and PW-DFT calculations used resources of the Argonne Leadership Computing Facility, which is a DOE Office of Science User Facility supported under contract DE-AC02-06CH11357. The DFT calculation using Gaussian orbitals were carried out on computing resources in the University of Pittsburgh's Center for Research Computing. K.D.J. acknowledges NSF (CBET-2028826) for partial support of this work. S.U. was supported in part by the Pittsburgh Quantum Institute (PQI) Graduate Quantum Leader Award.

\bibliography{theory}

\begin{thebibliography}{72}%
\makeatletter
\providecommand \@ifxundefined [1]{%
 \@ifx{#1\undefined}
}%
\providecommand \@ifnum [1]{%
 \ifnum #1\expandafter \@firstoftwo
 \else \expandafter \@secondoftwo
 \fi
}%
\providecommand \@ifx [1]{%
 \ifx #1\expandafter \@firstoftwo
 \else \expandafter \@secondoftwo
 \fi
}%
\providecommand \natexlab [1]{#1}%
\providecommand \enquote  [1]{``#1''}%
\providecommand \bibnamefont  [1]{#1}%
\providecommand \bibfnamefont [1]{#1}%
\providecommand \citenamefont [1]{#1}%
\providecommand \href@noop [0]{\@secondoftwo}%
\providecommand \href [0]{\begingroup \@sanitize@url \@href}%
\providecommand \@href[1]{\@@startlink{#1}\@@href}%
\providecommand \@@href[1]{\endgroup#1\@@endlink}%
\providecommand \@sanitize@url [0]{\catcode `\\12\catcode `\$12\catcode
  `\&12\catcode `\#12\catcode `\^12\catcode `\_12\catcode `\%12\relax}%
\providecommand \@@startlink[1]{}%
\providecommand \@@endlink[0]{}%
\providecommand \url  [0]{\begingroup\@sanitize@url \@url }%
\providecommand \@url [1]{\endgroup\@href {#1}{\urlprefix }}%
\providecommand \urlprefix  [0]{URL }%
\providecommand \Eprint [0]{\href }%
\providecommand \doibase [0]{http://dx.doi.org/}%
\providecommand \selectlanguage [0]{\@gobble}%
\providecommand \bibinfo  [0]{\@secondoftwo}%
\providecommand \bibfield  [0]{\@secondoftwo}%
\providecommand \translation [1]{[#1]}%
\providecommand \BibitemOpen [0]{}%
\providecommand \bibitemStop [0]{}%
\providecommand \bibitemNoStop [0]{.\EOS\space}%
\providecommand \EOS [0]{\spacefactor3000\relax}%
\providecommand \BibitemShut  [1]{\csname bibitem#1\endcsname}%
\let\auto@bib@innerbib\@empty
\bibitem [{\citenamefont {Hughes}, \citenamefont {Shaw},\ and\ \citenamefont
  {Russo}(2020)}]{doi:10.1021/acs.jpca.0c06595}%
  \BibitemOpen
  \bibfield  {author} {\bibinfo {author} {\bibfnamefont {T.~J.}\ \bibnamefont
  {Hughes}}, \bibinfo {author} {\bibfnamefont {R.~A.}\ \bibnamefont {Shaw}}, \
  and\ \bibinfo {author} {\bibfnamefont {S.~P.}\ \bibnamefont {Russo}},\
  }\bibfield  {title} {\enquote {\bibinfo {title} {{Computational
  Investigations of Dispersion Interactions between Small Molecules and
  Graphene-like Flakes}},}\ }\href {\doibase 10.1021/acs.jpca.0c06595}
  {\bibfield  {journal} {\bibinfo  {journal} {J. Phys. Chem. A}\ }\textbf
  {\bibinfo {volume} {124}},\ \bibinfo {pages} {9552--9561} (\bibinfo {year}
  {2020})}\BibitemShut {NoStop}%
\bibitem [{\citenamefont {Bartolomei}\ \emph {et~al.}(2021)\citenamefont
  {Bartolomei}, \citenamefont {Hernández}, \citenamefont {Campos-Martínez},
  \citenamefont {Hernández-Lamoneda},\ and\ \citenamefont
  {Giorgi}}]{10.1016/j.carbon.2021.02.056}%
  \BibitemOpen
  \bibfield  {author} {\bibinfo {author} {\bibfnamefont {M.}~\bibnamefont
  {Bartolomei}}, \bibinfo {author} {\bibfnamefont {M.~I.}\ \bibnamefont
  {Hernández}}, \bibinfo {author} {\bibfnamefont {J.}~\bibnamefont
  {Campos-Martínez}}, \bibinfo {author} {\bibfnamefont {R.}~\bibnamefont
  {Hernández-Lamoneda}}, \ and\ \bibinfo {author} {\bibfnamefont
  {G.}~\bibnamefont {Giorgi}},\ }\bibfield  {title} {\enquote {\bibinfo {title}
  {{Permeation of chemisorbed hydrogen through graphene: A flipping mechanism
  elucidated}},}\ }\href {\doibase 10.1016/j.carbon.2021.02.056} {\bibfield
  {journal} {\bibinfo  {journal} {Carbon}\ }\textbf {\bibinfo {volume} {178}},\
  \bibinfo {pages} {718--727} (\bibinfo {year} {2021})}\BibitemShut {NoStop}%
\bibitem [{\citenamefont {Ma}, \citenamefont {Michaelides},\ and\ \citenamefont
  {Alf{\`e}}(2011)}]{10.1063/1.3569134}%
  \BibitemOpen
  \bibfield  {author} {\bibinfo {author} {\bibfnamefont {J.}~\bibnamefont
  {Ma}}, \bibinfo {author} {\bibfnamefont {A.}~\bibnamefont {Michaelides}}, \
  and\ \bibinfo {author} {\bibfnamefont {D.}~\bibnamefont {Alf{\`e}}},\
  }\bibfield  {title} {\enquote {\bibinfo {title} {{Binding of hydrogen on
  benzene, coronene, and graphene from quantum {Monte} {Carlo}
  calculations}},}\ }\href {\doibase 10.1063/1.3569134} {\bibfield  {journal}
  {\bibinfo  {journal} {J. Chem. Phys.}\ }\textbf {\bibinfo {volume} {134}},\
  \bibinfo {pages} {134701} (\bibinfo {year} {2011})}\BibitemShut {NoStop}%
\bibitem [{\citenamefont {Shin}\ \emph
  {et~al.}(2019{\natexlab{a}})\citenamefont {Shin}, \citenamefont {Luo},
  \citenamefont {Benali},\ and\ \citenamefont {Kwon}}]{shin_diffusion_2019}%
  \BibitemOpen
  \bibfield  {author} {\bibinfo {author} {\bibfnamefont {H.}~\bibnamefont
  {Shin}}, \bibinfo {author} {\bibfnamefont {Y.}~\bibnamefont {Luo}}, \bibinfo
  {author} {\bibfnamefont {A.}~\bibnamefont {Benali}}, \ and\ \bibinfo {author}
  {\bibfnamefont {Y.}~\bibnamefont {Kwon}},\ }\bibfield  {title} {\enquote
  {\bibinfo {title} {{Diffusion {Monte} {Carlo} study of \ce{O2} adsorption on
  single layer graphene}},}\ }\href {\doibase 10.1103/PhysRevB.100.075430}
  {\bibfield  {journal} {\bibinfo  {journal} {Phys. Rev. B}\ }\textbf {\bibinfo
  {volume} {100}},\ \bibinfo {pages} {075430} (\bibinfo {year}
  {2019}{\natexlab{a}})}\BibitemShut {NoStop}%
\bibitem [{\citenamefont {Ma}\ \emph {et~al.}(2011)\citenamefont {Ma},
  \citenamefont {Michaelides}, \citenamefont {Alf\`e}, \citenamefont {Schimka},
  \citenamefont {Kresse},\ and\ \citenamefont
  {Wang}}]{10.1103/PhysRevB.84.033402}%
  \BibitemOpen
  \bibfield  {author} {\bibinfo {author} {\bibfnamefont {J.}~\bibnamefont
  {Ma}}, \bibinfo {author} {\bibfnamefont {A.}~\bibnamefont {Michaelides}},
  \bibinfo {author} {\bibfnamefont {D.}~\bibnamefont {Alf\`e}}, \bibinfo
  {author} {\bibfnamefont {L.}~\bibnamefont {Schimka}}, \bibinfo {author}
  {\bibfnamefont {G.}~\bibnamefont {Kresse}}, \ and\ \bibinfo {author}
  {\bibfnamefont {E.}~\bibnamefont {Wang}},\ }\bibfield  {title} {\enquote
  {\bibinfo {title} {{Adsorption and diffusion of water on graphene from first
  principles}},}\ }\href {\doibase 10.1103/PhysRevB.84.033402} {\bibfield
  {journal} {\bibinfo  {journal} {Phys. Rev. B}\ }\textbf {\bibinfo {volume}
  {84}},\ \bibinfo {pages} {033402} (\bibinfo {year} {2011})}\BibitemShut
  {NoStop}%
\bibitem [{\citenamefont {Ahn}\ \emph {et~al.}(2021)\citenamefont {Ahn},
  \citenamefont {Hong}, \citenamefont {Lee}, \citenamefont {Shin},
  \citenamefont {Benali},\ and\ \citenamefont {Kwon}}]{D1CP02473F}%
  \BibitemOpen
  \bibfield  {author} {\bibinfo {author} {\bibfnamefont {J.}~\bibnamefont
  {Ahn}}, \bibinfo {author} {\bibfnamefont {I.}~\bibnamefont {Hong}}, \bibinfo
  {author} {\bibfnamefont {G.}~\bibnamefont {Lee}}, \bibinfo {author}
  {\bibfnamefont {H.}~\bibnamefont {Shin}}, \bibinfo {author} {\bibfnamefont
  {A.}~\bibnamefont {Benali}}, \ and\ \bibinfo {author} {\bibfnamefont
  {Y.}~\bibnamefont {Kwon}},\ }\bibfield  {title} {\enquote {\bibinfo {title}
  {Adsorption of a single pt atom on graphene: spin crossing between
  physisorbed triplet and chemisorbed singlet states},}\ }\href {\doibase
  10.1039/D1CP02473F} {\bibfield  {journal} {\bibinfo  {journal} {Phys. Chem.
  Chem. Phys.}\ }\textbf {\bibinfo {volume} {23}},\ \bibinfo {pages}
  {22147--22154} (\bibinfo {year} {2021})}\BibitemShut {NoStop}%
\bibitem [{\citenamefont {Geim}(2009)}]{doi:10.1126/science.1158877}%
  \BibitemOpen
  \bibfield  {author} {\bibinfo {author} {\bibfnamefont {A.~K.}\ \bibnamefont
  {Geim}},\ }\bibfield  {title} {\enquote {\bibinfo {title} {{Graphene: Status
  and Prospects}},}\ }\href {\doibase 10.1126/science.1158877} {\bibfield
  {journal} {\bibinfo  {journal} {Science}\ }\textbf {\bibinfo {volume}
  {324}},\ \bibinfo {pages} {1530--1534} (\bibinfo {year} {2009})}\BibitemShut
  {NoStop}%
\bibitem [{\citenamefont {Novoselov}\ \emph {et~al.}(2005)\citenamefont
  {Novoselov}, \citenamefont {Geim}, \citenamefont {Morozov}, \citenamefont
  {Jiang}, \citenamefont {Katsnelson}, \citenamefont {Grigorieva},
  \citenamefont {Dubonos},\ and\ \citenamefont {Firsov}}]{10.1038/nature04233}%
  \BibitemOpen
  \bibfield  {author} {\bibinfo {author} {\bibfnamefont {K.~S.}\ \bibnamefont
  {Novoselov}}, \bibinfo {author} {\bibfnamefont {A.~K.}\ \bibnamefont {Geim}},
  \bibinfo {author} {\bibfnamefont {S.~V.}\ \bibnamefont {Morozov}}, \bibinfo
  {author} {\bibfnamefont {D.}~\bibnamefont {Jiang}}, \bibinfo {author}
  {\bibfnamefont {M.~I.}\ \bibnamefont {Katsnelson}}, \bibinfo {author}
  {\bibfnamefont {I.~V.}\ \bibnamefont {Grigorieva}}, \bibinfo {author}
  {\bibfnamefont {S.~V.}\ \bibnamefont {Dubonos}}, \ and\ \bibinfo {author}
  {\bibfnamefont {A.~A.}\ \bibnamefont {Firsov}},\ }\bibfield  {title}
  {\enquote {\bibinfo {title} {{Two-dimensional gas of massless Dirac fermions
  in graphene}},}\ }\href {\doibase 10.1038/nature04233} {\bibfield  {journal}
  {\bibinfo  {journal} {Nature}\ }\textbf {\bibinfo {volume} {438}},\ \bibinfo
  {pages} {197--200} (\bibinfo {year} {2005})}\BibitemShut {NoStop}%
\bibitem [{\citenamefont {Alekseeva}\ \emph {et~al.}(2020)\citenamefont
  {Alekseeva}, \citenamefont {Pushkareva}, \citenamefont {Pushkarev},\ and\
  \citenamefont {Fateev}}]{Alekseeva2020}%
  \BibitemOpen
  \bibfield  {author} {\bibinfo {author} {\bibfnamefont {O.~K.}\ \bibnamefont
  {Alekseeva}}, \bibinfo {author} {\bibfnamefont {I.~V.}\ \bibnamefont
  {Pushkareva}}, \bibinfo {author} {\bibfnamefont {A.~S.}\ \bibnamefont
  {Pushkarev}}, \ and\ \bibinfo {author} {\bibfnamefont {V.~N.}\ \bibnamefont
  {Fateev}},\ }\bibfield  {title} {\enquote {\bibinfo {title} {{Graphene and
  Graphene-Like Materials for Hydrogen Energy}},}\ }\href {\doibase
  10.1134/S1995078020030027} {\bibfield  {journal} {\bibinfo  {journal}
  {{Nanotechnol. Russ.}}\ }\textbf {\bibinfo {volume} {15}},\ \bibinfo {pages}
  {273--300} (\bibinfo {year} {2020})}\BibitemShut {NoStop}%
\bibitem [{\citenamefont {Niaz}, \citenamefont {Manzoor},\ and\ \citenamefont
  {Pandith}(2015)}]{10.1016/j.rser.2015.05.011}%
  \BibitemOpen
  \bibfield  {author} {\bibinfo {author} {\bibfnamefont {S.}~\bibnamefont
  {Niaz}}, \bibinfo {author} {\bibfnamefont {T.}~\bibnamefont {Manzoor}}, \
  and\ \bibinfo {author} {\bibfnamefont {A.~H.}\ \bibnamefont {Pandith}},\
  }\bibfield  {title} {\enquote {\bibinfo {title} {{Hydrogen storage:
  Materials, methods and perspectives}},}\ }\href {\doibase
  10.1016/j.rser.2015.05.011} {\bibfield  {journal} {\bibinfo  {journal}
  {Renew. Sustain. Energy Rev}\ }\textbf {\bibinfo {volume} {50}},\ \bibinfo
  {pages} {457--469} (\bibinfo {year} {2015})}\BibitemShut {NoStop}%
\bibitem [{\citenamefont {Fomkin}\ \emph {et~al.}(2021)\citenamefont {Fomkin},
  \citenamefont {Pribylov}, \citenamefont {Men’shchikov}, \citenamefont
  {Shkolin}, \citenamefont {Aksyutin}, \citenamefont {Ishkov}, \citenamefont
  {Romanov},\ and\ \citenamefont {Khozina}}]{reactions2030014}%
  \BibitemOpen
  \bibfield  {author} {\bibinfo {author} {\bibfnamefont {A.}~\bibnamefont
  {Fomkin}}, \bibinfo {author} {\bibfnamefont {A.}~\bibnamefont {Pribylov}},
  \bibinfo {author} {\bibfnamefont {I.}~\bibnamefont {Men’shchikov}},
  \bibinfo {author} {\bibfnamefont {A.}~\bibnamefont {Shkolin}}, \bibinfo
  {author} {\bibfnamefont {O.}~\bibnamefont {Aksyutin}}, \bibinfo {author}
  {\bibfnamefont {A.}~\bibnamefont {Ishkov}}, \bibinfo {author} {\bibfnamefont
  {K.}~\bibnamefont {Romanov}}, \ and\ \bibinfo {author} {\bibfnamefont
  {E.}~\bibnamefont {Khozina}},\ }\bibfield  {title} {\enquote {\bibinfo
  {title} {{Adsorption-Based Hydrogen Storage in Activated Carbons and Model
  Carbon Structures}},}\ }\href {\doibase 10.3390/reactions2030014} {\bibfield
  {journal} {\bibinfo  {journal} {Reactions}\ }\textbf {\bibinfo {volume}
  {2}},\ \bibinfo {pages} {209--226} (\bibinfo {year} {2021})}\BibitemShut
  {NoStop}%
\bibitem [{\citenamefont {Ataca}\ \emph {et~al.}(2008)\citenamefont {Ataca},
  \citenamefont {Aktürk}, \citenamefont {Ciraci},\ and\ \citenamefont
  {Ustunel}}]{doi:10.1063/1.2963976}%
  \BibitemOpen
  \bibfield  {author} {\bibinfo {author} {\bibfnamefont {C.}~\bibnamefont
  {Ataca}}, \bibinfo {author} {\bibfnamefont {E.}~\bibnamefont {Aktürk}},
  \bibinfo {author} {\bibfnamefont {S.}~\bibnamefont {Ciraci}}, \ and\ \bibinfo
  {author} {\bibfnamefont {H.}~\bibnamefont {Ustunel}},\ }\bibfield  {title}
  {\enquote {\bibinfo {title} {{High-capacity hydrogen storage by metallized
  graphene}},}\ }\href {\doibase 10.1063/1.2963976} {\bibfield  {journal}
  {\bibinfo  {journal} {Appl. Phys. Lett.}\ }\textbf {\bibinfo {volume} {93}},\
  \bibinfo {pages} {043123} (\bibinfo {year} {2008})}\BibitemShut {NoStop}%
\bibitem [{\citenamefont {Dimitrakakis}, \citenamefont {Tylianakis},\ and\
  \citenamefont {Froudakis}(2008)}]{doi:10.1021/nl801417w}%
  \BibitemOpen
  \bibfield  {author} {\bibinfo {author} {\bibfnamefont {G.~K.}\ \bibnamefont
  {Dimitrakakis}}, \bibinfo {author} {\bibfnamefont {E.}~\bibnamefont
  {Tylianakis}}, \ and\ \bibinfo {author} {\bibfnamefont {G.~E.}\ \bibnamefont
  {Froudakis}},\ }\bibfield  {title} {\enquote {\bibinfo {title} {{Pillared
  Graphene: A New 3-D Network Nanostructure for Enhanced Hydrogen Storage}},}\
  }\href {\doibase 10.1021/nl801417w} {\bibfield  {journal} {\bibinfo
  {journal} {Nano Lett.}\ }\textbf {\bibinfo {volume} {8}},\ \bibinfo {pages}
  {3166--3170} (\bibinfo {year} {2008})}\BibitemShut {NoStop}%
\bibitem [{\citenamefont {Ferullo}, \citenamefont {Domancich},\ and\
  \citenamefont
  {Castellani}(2010)}]{doi:10.1016/j.cplett.2010.10.1002/wcms.136010.027}%
  \BibitemOpen
  \bibfield  {author} {\bibinfo {author} {\bibfnamefont {R.~M.}\ \bibnamefont
  {Ferullo}}, \bibinfo {author} {\bibfnamefont {N.~F.}\ \bibnamefont
  {Domancich}}, \ and\ \bibinfo {author} {\bibfnamefont {N.~J.}\ \bibnamefont
  {Castellani}},\ }\bibfield  {title} {\enquote {\bibinfo {title} {{On the
  performance of van der Waals corrected-density functional theory in
  describing the atomic hydrogen physisorption on graphite}},}\ }\href
  {\doibase 10.1016/j.cplett.2010.10.027} {\bibfield  {journal} {\bibinfo
  {journal} {Chem. Phys. Lett}\ }\textbf {\bibinfo {volume} {500}},\ \bibinfo
  {pages} {283--286} (\bibinfo {year} {2010})}\BibitemShut {NoStop}%
\bibitem [{\citenamefont {Sha}\ and\ \citenamefont
  {Jackson}(2002)}]{SHA2002318}%
  \BibitemOpen
  \bibfield  {author} {\bibinfo {author} {\bibfnamefont {X.}~\bibnamefont
  {Sha}}\ and\ \bibinfo {author} {\bibfnamefont {B.}~\bibnamefont {Jackson}},\
  }\bibfield  {title} {\enquote {\bibinfo {title} {{First-principles study of
  the structural and energetic properties of H atoms on a graphite (0001)
  surface}},}\ }\href {\doibase 10.1016/S0039-6028(01)01602-8} {\bibfield
  {journal} {\bibinfo  {journal} {Surf. Sci.}\ }\textbf {\bibinfo {volume}
  {496}},\ \bibinfo {pages} {318--330} (\bibinfo {year} {2002})}\BibitemShut
  {NoStop}%
\bibitem [{\citenamefont {Jeloaica}\ and\ \citenamefont
  {Sidis}(1999)}]{JELOAICA1999157}%
  \BibitemOpen
  \bibfield  {author} {\bibinfo {author} {\bibfnamefont {L.}~\bibnamefont
  {Jeloaica}}\ and\ \bibinfo {author} {\bibfnamefont {V.}~\bibnamefont
  {Sidis}},\ }\bibfield  {title} {\enquote {\bibinfo {title} {{DFT
  investigation of the adsorption of atomic hydrogen on a cluster-model
  graphite surface}},}\ }\href {\doibase 10.1016/S0009-2614(98)01337-2}
  {\bibfield  {journal} {\bibinfo  {journal} {Chem. Phys. Lett.}\ }\textbf
  {\bibinfo {volume} {300}},\ \bibinfo {pages} {157--162} (\bibinfo {year}
  {1999})}\BibitemShut {NoStop}%
\bibitem [{\citenamefont {Sahu}\ and\ \citenamefont
  {Rout}(2017)}]{10.1007/s40089-017-0203-5}%
  \BibitemOpen
  \bibfield  {author} {\bibinfo {author} {\bibfnamefont {S.}~\bibnamefont
  {Sahu}}\ and\ \bibinfo {author} {\bibfnamefont {G.~C.}\ \bibnamefont
  {Rout}},\ }\bibfield  {title} {\enquote {\bibinfo {title} {{Band gap opening
  in graphene: a short theoretical study}},}\ }\href {\doibase
  10.1007/s40089-017-0203-5} {\bibfield  {journal} {\bibinfo  {journal} {Int.
  Nano Lett.}\ }\textbf {\bibinfo {volume} {7}},\ \bibinfo {pages} {81--89}
  (\bibinfo {year} {2017})}\BibitemShut {NoStop}%
\bibitem [{\citenamefont {Gonz\'{a}lez-Herrero}\ \emph
  {et~al.}(2016)\citenamefont {Gonz\'{a}lez-Herrero}, \citenamefont
  {G\'{o}mez-Rodr\'{i}guez}, \citenamefont {Mallet}, \citenamefont {Moaied},
  \citenamefont {Palacios}, \citenamefont {Salgado}, \citenamefont {Ugeda},
  \citenamefont {Veuillen}, \citenamefont {Yndurain},\ and\ \citenamefont
  {Brihuega}}]{gonzalez-herrero_atomic-scale_2016}%
  \BibitemOpen
  \bibfield  {author} {\bibinfo {author} {\bibfnamefont {H.}~\bibnamefont
  {Gonz\'{a}lez-Herrero}}, \bibinfo {author} {\bibfnamefont {J.~M.}\
  \bibnamefont {G\'{o}mez-Rodr\'{i}guez}}, \bibinfo {author} {\bibfnamefont
  {P.}~\bibnamefont {Mallet}}, \bibinfo {author} {\bibfnamefont
  {M.}~\bibnamefont {Moaied}}, \bibinfo {author} {\bibfnamefont {J.~J.}\
  \bibnamefont {Palacios}}, \bibinfo {author} {\bibfnamefont {C.}~\bibnamefont
  {Salgado}}, \bibinfo {author} {\bibfnamefont {M.~M.}\ \bibnamefont {Ugeda}},
  \bibinfo {author} {\bibfnamefont {J.-Y.}\ \bibnamefont {Veuillen}}, \bibinfo
  {author} {\bibfnamefont {F.}~\bibnamefont {Yndurain}}, \ and\ \bibinfo
  {author} {\bibfnamefont {I.}~\bibnamefont {Brihuega}},\ }\bibfield  {title}
  {\enquote {\bibinfo {title} {{Atomic-scale control of graphene magnetism by
  using hydrogen atoms}},}\ }\href {\doibase 10.1126/science.aad8038}
  {\bibfield  {journal} {\bibinfo  {journal} {Science}\ }\textbf {\bibinfo
  {volume} {352}},\ \bibinfo {pages} {437--441} (\bibinfo {year}
  {2016})}\BibitemShut {NoStop}%
\bibitem [{\citenamefont {Gonz{\'{a}}lez-Herrero}\ \emph
  {et~al.}(2019{\natexlab{a}})\citenamefont {Gonz{\'{a}}lez-Herrero},
  \citenamefont {del R{\'{\i}}o}, \citenamefont {Mallet}, \citenamefont
  {Veuillen}, \citenamefont {Palacios}, \citenamefont
  {G{\'{o}}mez-Rodr{\'{\i}}guez}, \citenamefont {Brihuega},\ and\ \citenamefont
  {Yndur{\'{a}}in}}]{GonzlezHerrero2019}%
  \BibitemOpen
  \bibfield  {author} {\bibinfo {author} {\bibfnamefont {H.}~\bibnamefont
  {Gonz{\'{a}}lez-Herrero}}, \bibinfo {author} {\bibfnamefont {E.~C.}\
  \bibnamefont {del R{\'{\i}}o}}, \bibinfo {author} {\bibfnamefont
  {P.}~\bibnamefont {Mallet}}, \bibinfo {author} {\bibfnamefont {J.-Y.}\
  \bibnamefont {Veuillen}}, \bibinfo {author} {\bibfnamefont {J.~J.}\
  \bibnamefont {Palacios}}, \bibinfo {author} {\bibfnamefont {J.~M.}\
  \bibnamefont {G{\'{o}}mez-Rodr{\'{\i}}guez}}, \bibinfo {author}
  {\bibfnamefont {I.}~\bibnamefont {Brihuega}}, \ and\ \bibinfo {author}
  {\bibfnamefont {F.}~\bibnamefont {Yndur{\'{a}}in}},\ }\bibfield  {title}
  {\enquote {\bibinfo {title} {{Hydrogen physisorption channel on graphene: a
  highway for atomic H diffusion}},}\ }\href {\doibase
  10.1088/2053-1583/ab03a0} {\bibfield  {journal} {\bibinfo  {journal} {2D
  Mater.}\ }\textbf {\bibinfo {volume} {6}},\ \bibinfo {pages} {021004}
  (\bibinfo {year} {2019}{\natexlab{a}})}\BibitemShut {NoStop}%
\bibitem [{\citenamefont {Gonz{\'{a}}lez-Herrero}\ \emph
  {et~al.}(2019{\natexlab{b}})\citenamefont {Gonz{\'{a}}lez-Herrero},
  \citenamefont {del R{\'{i}}o}, \citenamefont {Mallet}, \citenamefont
  {Veuillen}, \citenamefont {Palacios}, \citenamefont
  {G{\'{o}}mez-Rodr{\'{i}}guez}, \citenamefont {Brihuega},\ and\ \citenamefont
  {Yndur{\'{a}}in}}]{10.1088/2053-1583/ab03a0}%
  \BibitemOpen
  \bibfield  {author} {\bibinfo {author} {\bibfnamefont {H.}~\bibnamefont
  {Gonz{\'{a}}lez-Herrero}}, \bibinfo {author} {\bibfnamefont {E.~C.}\
  \bibnamefont {del R{\'{i}}o}}, \bibinfo {author} {\bibfnamefont
  {P.}~\bibnamefont {Mallet}}, \bibinfo {author} {\bibfnamefont {J.-Y.}\
  \bibnamefont {Veuillen}}, \bibinfo {author} {\bibfnamefont {J.~J.}\
  \bibnamefont {Palacios}}, \bibinfo {author} {\bibfnamefont {J.~M.}\
  \bibnamefont {G{\'{o}}mez-Rodr{\'{i}}guez}}, \bibinfo {author} {\bibfnamefont
  {I.}~\bibnamefont {Brihuega}}, \ and\ \bibinfo {author} {\bibfnamefont
  {F.}~\bibnamefont {Yndur{\'{a}}in}},\ }\bibfield  {title} {\enquote {\bibinfo
  {title} {{Hydrogen physisorption channel on graphene: a highway for atomic H
  diffusion}},}\ }\href {\doibase 10.1088/2053-1583/ab03a0} {\bibfield
  {journal} {\bibinfo  {journal} {2d Mater.}\ }\textbf {\bibinfo {volume}
  {6}},\ \bibinfo {pages} {021004} (\bibinfo {year}
  {2019}{\natexlab{b}})}\BibitemShut {NoStop}%
\bibitem [{\citenamefont {Hummel}, \citenamefont {Tsatsoulis},\ and\
  \citenamefont {Gr{\"u}neis}(2017)}]{doi:10.1063/1.4977994}%
  \BibitemOpen
  \bibfield  {author} {\bibinfo {author} {\bibfnamefont {F.}~\bibnamefont
  {Hummel}}, \bibinfo {author} {\bibfnamefont {T.}~\bibnamefont {Tsatsoulis}},
  \ and\ \bibinfo {author} {\bibfnamefont {A.}~\bibnamefont {Gr{\"u}neis}},\
  }\bibfield  {title} {\enquote {\bibinfo {title} {{Low rank factorization of
  the Coulomb integrals for periodic coupled cluster theory}},}\ }\href
  {\doibase 10.1063/1.4977994} {\bibfield  {journal} {\bibinfo  {journal} {J.
  Chem. Phys.}\ }\textbf {\bibinfo {volume} {146}},\ \bibinfo {pages} {124105}
  (\bibinfo {year} {2017})}\BibitemShut {NoStop}%
\bibitem [{\citenamefont {Mihm}, \citenamefont {McIsaac},\ and\ \citenamefont
  {Shepherd}(2019)}]{doi:10.1063/1.5091445}%
  \BibitemOpen
  \bibfield  {author} {\bibinfo {author} {\bibfnamefont {T.~N.}\ \bibnamefont
  {Mihm}}, \bibinfo {author} {\bibfnamefont {A.~R.}\ \bibnamefont {McIsaac}}, \
  and\ \bibinfo {author} {\bibfnamefont {J.~J.}\ \bibnamefont {Shepherd}},\
  }\bibfield  {title} {\enquote {\bibinfo {title} {{An optimized twist angle to
  find the twist-averaged correlation energy applied to the uniform electron
  gas}},}\ }\href {\doibase 10.1063/1.5091445} {\bibfield  {journal} {\bibinfo
  {journal} {J. Chem. Phys.}\ }\textbf {\bibinfo {volume} {150}},\ \bibinfo
  {pages} {191101} (\bibinfo {year} {2019})}\BibitemShut {NoStop}%
\bibitem [{\citenamefont {Callahan}, \citenamefont {Lange},\ and\ \citenamefont
  {Berkelbach}(2021)}]{doi:10.1063/5.0049890}%
  \BibitemOpen
  \bibfield  {author} {\bibinfo {author} {\bibfnamefont {J.~M.}\ \bibnamefont
  {Callahan}}, \bibinfo {author} {\bibfnamefont {M.~F.}\ \bibnamefont {Lange}},
  \ and\ \bibinfo {author} {\bibfnamefont {T.~C.}\ \bibnamefont {Berkelbach}},\
  }\bibfield  {title} {\enquote {\bibinfo {title} {{Dynamical correlation
  energy of metals in large basis sets from downfolding and composite
  approaches}},}\ }\href {\doibase 10.1063/5.0049890} {\bibfield  {journal}
  {\bibinfo  {journal} {J. Chem. Phys.}\ }\textbf {\bibinfo {volume} {154}},\
  \bibinfo {pages} {211105} (\bibinfo {year} {2021})}\BibitemShut {NoStop}%
\bibitem [{\citenamefont {Booth}\ \emph {et~al.}(2013)\citenamefont {Booth},
  \citenamefont {Gr{\"u}neis}, \citenamefont {Kresse},\ and\ \citenamefont
  {Alavi}}]{10.1038/nature11770}%
  \BibitemOpen
  \bibfield  {author} {\bibinfo {author} {\bibfnamefont {G.~H.}\ \bibnamefont
  {Booth}}, \bibinfo {author} {\bibfnamefont {A.}~\bibnamefont {Gr{\"u}neis}},
  \bibinfo {author} {\bibfnamefont {G.}~\bibnamefont {Kresse}}, \ and\ \bibinfo
  {author} {\bibfnamefont {A.}~\bibnamefont {Alavi}},\ }\bibfield  {title}
  {\enquote {\bibinfo {title} {{Towards an exact description of electronic
  wavefunctions in real solids}},}\ }\href {\doibase 10.1038/nature11770}
  {\bibfield  {journal} {\bibinfo  {journal} {Nature}\ }\textbf {\bibinfo
  {volume} {493}},\ \bibinfo {pages} {365--370} (\bibinfo {year}
  {2013})}\BibitemShut {NoStop}%
\bibitem [{\citenamefont {Sch{\"a}fer}\ \emph {et~al.}(2021)\citenamefont
  {Sch{\"a}fer}, \citenamefont {Libisch}, \citenamefont {Kresse},\ and\
  \citenamefont {Gr{\"u}neis}}]{doi:10.1063/5.0036363}%
  \BibitemOpen
  \bibfield  {author} {\bibinfo {author} {\bibfnamefont {T.}~\bibnamefont
  {Sch{\"a}fer}}, \bibinfo {author} {\bibfnamefont {F.}~\bibnamefont
  {Libisch}}, \bibinfo {author} {\bibfnamefont {G.}~\bibnamefont {Kresse}}, \
  and\ \bibinfo {author} {\bibfnamefont {A.}~\bibnamefont {Gr{\"u}neis}},\
  }\bibfield  {title} {\enquote {\bibinfo {title} {{Local embedding of coupled
  cluster theory into the random phase approximation using plane waves}},}\
  }\href {\doibase 10.1063/5.0036363} {\bibfield  {journal} {\bibinfo
  {journal} {J. Chem. Phys.}\ }\textbf {\bibinfo {volume} {154}},\ \bibinfo
  {pages} {011101} (\bibinfo {year} {2021})}\BibitemShut {NoStop}%
\bibitem [{\citenamefont {Benali}\ \emph {et~al.}(2020)\citenamefont {Benali},
  \citenamefont {Gasperich}, \citenamefont {Jordan}, \citenamefont
  {Applencourt}, \citenamefont {Luo}, \citenamefont {Bennett}, \citenamefont
  {Krogel}, \citenamefont {Shulenburger}, \citenamefont {Kent}, \citenamefont
  {Loos}, \citenamefont {Scemama},\ and\ \citenamefont
  {Caffarel}}]{doi:10.1063/5.0021036}%
  \BibitemOpen
  \bibfield  {author} {\bibinfo {author} {\bibfnamefont {A.}~\bibnamefont
  {Benali}}, \bibinfo {author} {\bibfnamefont {K.}~\bibnamefont {Gasperich}},
  \bibinfo {author} {\bibfnamefont {K.~D.}\ \bibnamefont {Jordan}}, \bibinfo
  {author} {\bibfnamefont {T.}~\bibnamefont {Applencourt}}, \bibinfo {author}
  {\bibfnamefont {Y.}~\bibnamefont {Luo}}, \bibinfo {author} {\bibfnamefont
  {M.~C.}\ \bibnamefont {Bennett}}, \bibinfo {author} {\bibfnamefont {J.~T.}\
  \bibnamefont {Krogel}}, \bibinfo {author} {\bibfnamefont {L.}~\bibnamefont
  {Shulenburger}}, \bibinfo {author} {\bibfnamefont {P.~R.~C.}\ \bibnamefont
  {Kent}}, \bibinfo {author} {\bibfnamefont {P.-F.}\ \bibnamefont {Loos}},
  \bibinfo {author} {\bibfnamefont {A.}~\bibnamefont {Scemama}}, \ and\
  \bibinfo {author} {\bibfnamefont {M.}~\bibnamefont {Caffarel}},\ }\bibfield
  {title} {\enquote {\bibinfo {title} {{Toward a systematic improvement of the
  fixed-node approximation in diffusion Monte Carlo for solids—A case study
  in diamond}},}\ }\href {\doibase 10.1063/5.0021036} {\bibfield  {journal}
  {\bibinfo  {journal} {J. Chem. Phys.}\ }\textbf {\bibinfo {volume} {153}},\
  \bibinfo {pages} {184111} (\bibinfo {year} {2020})}\BibitemShut {NoStop}%
\bibitem [{\citenamefont {Sch{\"a}fer}, \citenamefont {Ramberger},\ and\
  \citenamefont {Kresse}(2017)}]{doi:10.1063/1.4976937}%
  \BibitemOpen
  \bibfield  {author} {\bibinfo {author} {\bibfnamefont {T.}~\bibnamefont
  {Sch{\"a}fer}}, \bibinfo {author} {\bibfnamefont {B.}~\bibnamefont
  {Ramberger}}, \ and\ \bibinfo {author} {\bibfnamefont {G.}~\bibnamefont
  {Kresse}},\ }\bibfield  {title} {\enquote {\bibinfo {title} {Quartic scaling
  mp2 for solids: A highly parallelized algorithm in the plane wave basis},}\
  }\href {\doibase 10.1063/1.4976937} {\bibfield  {journal} {\bibinfo
  {journal} {J. Chem. Phys.}\ }\textbf {\bibinfo {volume} {146}},\ \bibinfo
  {pages} {104101} (\bibinfo {year} {2017})}\BibitemShut {NoStop}%
\bibitem [{\citenamefont {Foulkes}\ \emph {et~al.}(2001)\citenamefont
  {Foulkes}, \citenamefont {Mitas}, \citenamefont {Needs},\ and\ \citenamefont
  {Rajagopal}}]{foulkes01}%
  \BibitemOpen
  \bibfield  {author} {\bibinfo {author} {\bibfnamefont {W.~M.~C.}\
  \bibnamefont {Foulkes}}, \bibinfo {author} {\bibfnamefont {L.}~\bibnamefont
  {Mitas}}, \bibinfo {author} {\bibfnamefont {R.~J.}\ \bibnamefont {Needs}}, \
  and\ \bibinfo {author} {\bibfnamefont {G.}~\bibnamefont {Rajagopal}},\
  }\bibfield  {title} {\enquote {\bibinfo {title} {Quantum monte carlo
  simulations of solids},}\ }\href@noop {} {\bibfield  {journal} {\bibinfo
  {journal} {Rev. Mod. Phys.}\ }\textbf {\bibinfo {volume} {73}},\ \bibinfo
  {pages} {33--83} (\bibinfo {year} {2001})}\BibitemShut {NoStop}%
\bibitem [{\citenamefont {Brandenburg}\ \emph {et~al.}(2019)\citenamefont
  {Brandenburg}, \citenamefont {Zen}, \citenamefont {Fitzner}, \citenamefont
  {Ramberger}, \citenamefont {Kresse}, \citenamefont {Tsatsoulis},
  \citenamefont {Gr{\"u}neis}, \citenamefont {Michaelides},\ and\ \citenamefont
  {Alf{\`e}}}]{doi:10.1021/acs.jpclett.8b03679}%
  \BibitemOpen
  \bibfield  {author} {\bibinfo {author} {\bibfnamefont {J.~G.}\ \bibnamefont
  {Brandenburg}}, \bibinfo {author} {\bibfnamefont {A.}~\bibnamefont {Zen}},
  \bibinfo {author} {\bibfnamefont {M.}~\bibnamefont {Fitzner}}, \bibinfo
  {author} {\bibfnamefont {B.}~\bibnamefont {Ramberger}}, \bibinfo {author}
  {\bibfnamefont {G.}~\bibnamefont {Kresse}}, \bibinfo {author} {\bibfnamefont
  {T.}~\bibnamefont {Tsatsoulis}}, \bibinfo {author} {\bibfnamefont
  {A.}~\bibnamefont {Gr{\"u}neis}}, \bibinfo {author} {\bibfnamefont
  {A.}~\bibnamefont {Michaelides}}, \ and\ \bibinfo {author} {\bibfnamefont
  {D.}~\bibnamefont {Alf{\`e}}},\ }\bibfield  {title} {\enquote {\bibinfo
  {title} {Physisorption of water on graphene: Subchemical accuracy from
  many-body electronic structure methods},}\ }\href {\doibase
  10.1021/acs.jpclett.8b03679} {\bibfield  {journal} {\bibinfo  {journal} {J.
  Phys. Chem. Lett.}\ }\textbf {\bibinfo {volume} {10}},\ \bibinfo {pages}
  {358--368} (\bibinfo {year} {2019})}\BibitemShut {NoStop}%
\bibitem [{\citenamefont {Perdew}, \citenamefont {Burke},\ and\ \citenamefont
  {Ernzerhof}(1996)}]{10.1103/PhysRevLett.77.3865}%
  \BibitemOpen
  \bibfield  {author} {\bibinfo {author} {\bibfnamefont {J.~P.}\ \bibnamefont
  {Perdew}}, \bibinfo {author} {\bibfnamefont {K.}~\bibnamefont {Burke}}, \
  and\ \bibinfo {author} {\bibfnamefont {M.}~\bibnamefont {Ernzerhof}},\
  }\bibfield  {title} {\enquote {\bibinfo {title} {{Generalized Gradient
  Approximation Made Simple}},}\ }\href {\doibase 10.1103/PhysRevLett.77.3865}
  {\bibfield  {journal} {\bibinfo  {journal} {Phys. Rev. Lett.}\ }\textbf
  {\bibinfo {volume} {77}},\ \bibinfo {pages} {3865--3868} (\bibinfo {year}
  {1996})}\BibitemShut {NoStop}%
\bibitem [{\citenamefont {Perdew}(1991)}]{PW91}%
  \BibitemOpen
  \bibfield  {author} {\bibinfo {author} {\bibfnamefont {J.~P.}\ \bibnamefont
  {Perdew}},\ }\bibfield  {title} {\enquote {\bibinfo {title} {{Unified theory
  of exchange and correlation beyond the local density approximation}},}\ }in\
  \href@noop {} {\emph {\bibinfo {booktitle} {Electronic Structure of Solids
  '91}}},\ \bibinfo {series} {Physical Research}, Vol.~\bibinfo {volume} {17},\
  \bibinfo {editor} {edited by\ \bibinfo {editor} {\bibfnamefont
  {P.}~\bibnamefont {Ziesche}}\ and\ \bibinfo {editor} {\bibfnamefont
  {H.}~\bibnamefont {Eschrig}}}\ (\bibinfo  {publisher} {Akademie Verlag},\
  \bibinfo {address} {Berlin},\ \bibinfo {year} {1991})\ pp.\ \bibinfo {pages}
  {11--20}\BibitemShut {NoStop}%
\bibitem [{\citenamefont {Roman}\ \emph {et~al.}(2007)\citenamefont {Roman},
  \citenamefont {Diño}, \citenamefont {Nakanishi}, \citenamefont {Kasai},
  \citenamefont {Sugimoto},\ and\ \citenamefont
  {Tange}}]{10.1016/j.carbon.2006.09.027}%
  \BibitemOpen
  \bibfield  {author} {\bibinfo {author} {\bibfnamefont {T.}~\bibnamefont
  {Roman}}, \bibinfo {author} {\bibfnamefont {W.~A.}\ \bibnamefont {Diño}},
  \bibinfo {author} {\bibfnamefont {H.}~\bibnamefont {Nakanishi}}, \bibinfo
  {author} {\bibfnamefont {H.}~\bibnamefont {Kasai}}, \bibinfo {author}
  {\bibfnamefont {T.}~\bibnamefont {Sugimoto}}, \ and\ \bibinfo {author}
  {\bibfnamefont {K.}~\bibnamefont {Tange}},\ }\bibfield  {title} {\enquote
  {\bibinfo {title} {Hydrogen pairing on graphene},}\ }\href {\doibase
  10.1016/j.carbon.2006.09.027} {\bibfield  {journal} {\bibinfo  {journal}
  {Carbon}\ }\textbf {\bibinfo {volume} {45}},\ \bibinfo {pages} {218--220}
  (\bibinfo {year} {2007})}\BibitemShut {NoStop}%
\bibitem [{\citenamefont {\u{S}ljivan\u{c}anin}\ \emph
  {et~al.}(2009)\citenamefont {\u{S}ljivan\u{c}anin}, \citenamefont {Rauls},
  \citenamefont {Hornek{\ae}r}, \citenamefont {Xu}, \citenamefont
  {Besenbacher},\ and\ \citenamefont {Hammer}}]{doi:10.1063/1.3187941}%
  \BibitemOpen
  \bibfield  {author} {\bibinfo {author} {\bibfnamefont {{\u{Z}}.}~\bibnamefont
  {\u{S}ljivan\u{c}anin}}, \bibinfo {author} {\bibfnamefont {E.}~\bibnamefont
  {Rauls}}, \bibinfo {author} {\bibfnamefont {L.}~\bibnamefont {Hornek{\ae}r}},
  \bibinfo {author} {\bibfnamefont {W.}~\bibnamefont {Xu}}, \bibinfo {author}
  {\bibfnamefont {F.}~\bibnamefont {Besenbacher}}, \ and\ \bibinfo {author}
  {\bibfnamefont {B.}~\bibnamefont {Hammer}},\ }\bibfield  {title} {\enquote
  {\bibinfo {title} {{Extended atomic hydrogen dimer configurations on the
  graphite(0001) surface}},}\ }\href {\doibase 10.1063/1.3187941} {\bibfield
  {journal} {\bibinfo  {journal} {J. Chem. Phys.}\ }\textbf {\bibinfo {volume}
  {131}},\ \bibinfo {pages} {084706} (\bibinfo {year} {2009})}\BibitemShut
  {NoStop}%
\bibitem [{\citenamefont {Lehtinen}\ \emph {et~al.}(2004)\citenamefont
  {Lehtinen}, \citenamefont {Foster}, \citenamefont {Ma}, \citenamefont
  {Krasheninnikov},\ and\ \citenamefont
  {Nieminen}}]{10.1103/PhysRevLett.93.187202}%
  \BibitemOpen
  \bibfield  {author} {\bibinfo {author} {\bibfnamefont {P.~O.}\ \bibnamefont
  {Lehtinen}}, \bibinfo {author} {\bibfnamefont {A.~S.}\ \bibnamefont
  {Foster}}, \bibinfo {author} {\bibfnamefont {Y.}~\bibnamefont {Ma}}, \bibinfo
  {author} {\bibfnamefont {A.~V.}\ \bibnamefont {Krasheninnikov}}, \ and\
  \bibinfo {author} {\bibfnamefont {R.~M.}\ \bibnamefont {Nieminen}},\
  }\bibfield  {title} {\enquote {\bibinfo {title} {Irradiation-induced
  magnetism in graphite: A density functional study},}\ }\href {\doibase
  10.1103/PhysRevLett.93.187202} {\bibfield  {journal} {\bibinfo  {journal}
  {Phys. Rev. Lett.}\ }\textbf {\bibinfo {volume} {93}},\ \bibinfo {pages}
  {187202} (\bibinfo {year} {2004})}\BibitemShut {NoStop}%
\bibitem [{\citenamefont {Lin}, \citenamefont {Ding},\ and\ \citenamefont
  {Yakobson}(2008)}]{10.1103/PhysRevB.78.041402}%
  \BibitemOpen
  \bibfield  {author} {\bibinfo {author} {\bibfnamefont {Y.}~\bibnamefont
  {Lin}}, \bibinfo {author} {\bibfnamefont {F.}~\bibnamefont {Ding}}, \ and\
  \bibinfo {author} {\bibfnamefont {B.~I.}\ \bibnamefont {Yakobson}},\
  }\bibfield  {title} {\enquote {\bibinfo {title} {{Hydrogen storage by
  spillover on graphene as a phase nucleation process}},}\ }\href {\doibase
  10.1103/PhysRevB.78.041402} {\bibfield  {journal} {\bibinfo  {journal} {Phys.
  Rev. B}\ }\textbf {\bibinfo {volume} {78}},\ \bibinfo {pages} {041402}
  (\bibinfo {year} {2008})}\BibitemShut {NoStop}%
\bibitem [{\citenamefont {Casolo}\ \emph {et~al.}(2009)\citenamefont {Casolo},
  \citenamefont {Løvvik}, \citenamefont {Martinazzo},\ and\ \citenamefont
  {Tantardini}}]{doi:10.1063/1.3072333}%
  \BibitemOpen
  \bibfield  {author} {\bibinfo {author} {\bibfnamefont {S.}~\bibnamefont
  {Casolo}}, \bibinfo {author} {\bibfnamefont {O.~M.}\ \bibnamefont {Løvvik}},
  \bibinfo {author} {\bibfnamefont {R.}~\bibnamefont {Martinazzo}}, \ and\
  \bibinfo {author} {\bibfnamefont {G.~F.}\ \bibnamefont {Tantardini}},\
  }\bibfield  {title} {\enquote {\bibinfo {title} {{Understanding adsorption of
  hydrogen atoms on graphene}},}\ }\href {\doibase 10.1063/1.3072333}
  {\bibfield  {journal} {\bibinfo  {journal} {J. Chem. Phys.}\ }\textbf
  {\bibinfo {volume} {130}},\ \bibinfo {pages} {054704} (\bibinfo {year}
  {2009})}\BibitemShut {NoStop}%
\bibitem [{\citenamefont {Miwa}, \citenamefont {Martins},\ and\ \citenamefont
  {Fazzio}(2008)}]{10.1088/0957-4484/19/15/155708}%
  \BibitemOpen
  \bibfield  {author} {\bibinfo {author} {\bibfnamefont {R.~H.}\ \bibnamefont
  {Miwa}}, \bibinfo {author} {\bibfnamefont {T.~B.}\ \bibnamefont {Martins}}, \
  and\ \bibinfo {author} {\bibfnamefont {A.}~\bibnamefont {Fazzio}},\
  }\bibfield  {title} {\enquote {\bibinfo {title} {Hydrogen adsorption on boron
  doped graphene: an ab initio study},}\ }\href {\doibase
  10.1088/0957-4484/19/15/155708} {\bibfield  {journal} {\bibinfo  {journal}
  {Nanotechnology}\ }\textbf {\bibinfo {volume} {19}},\ \bibinfo {pages}
  {155708} (\bibinfo {year} {2008})}\BibitemShut {NoStop}%
\bibitem [{\citenamefont {Ishii}\ \emph {et~al.}(2008)\citenamefont {Ishii},
  \citenamefont {Yamamoto}, \citenamefont {Asano},\ and\ \citenamefont
  {Fujiwara}}]{10.1088/1742-6596/100/5/052087}%
  \BibitemOpen
  \bibfield  {author} {\bibinfo {author} {\bibfnamefont {A.}~\bibnamefont
  {Ishii}}, \bibinfo {author} {\bibfnamefont {M.}~\bibnamefont {Yamamoto}},
  \bibinfo {author} {\bibfnamefont {H.}~\bibnamefont {Asano}}, \ and\ \bibinfo
  {author} {\bibfnamefont {K.}~\bibnamefont {Fujiwara}},\ }\bibfield  {title}
  {\enquote {\bibinfo {title} {{DFT} calculation for adatom adsorption on
  graphene sheet as a prototype of carbon nanotube functionalization},}\ }\href
  {\doibase 10.1088/1742-6596/100/5/052087} {\bibfield  {journal} {\bibinfo
  {journal} {J. Phys. Conf. Ser.}\ }\textbf {\bibinfo {volume} {100}},\
  \bibinfo {pages} {052087} (\bibinfo {year} {2008})}\BibitemShut {NoStop}%
\bibitem [{\citenamefont {Li}\ \emph {et~al.}(2010)\citenamefont {Li},
  \citenamefont {Zhao}, \citenamefont {He}, \citenamefont {Song}, \citenamefont
  {Lin}, \citenamefont {Liu}, \citenamefont {Xia},\ and\ \citenamefont
  {Mei}}]{10.1016/j.jmmm.2009.11.014}%
  \BibitemOpen
  \bibfield  {author} {\bibinfo {author} {\bibfnamefont {W.}~\bibnamefont
  {Li}}, \bibinfo {author} {\bibfnamefont {M.}~\bibnamefont {Zhao}}, \bibinfo
  {author} {\bibfnamefont {T.}~\bibnamefont {He}}, \bibinfo {author}
  {\bibfnamefont {C.}~\bibnamefont {Song}}, \bibinfo {author} {\bibfnamefont
  {X.}~\bibnamefont {Lin}}, \bibinfo {author} {\bibfnamefont {X.}~\bibnamefont
  {Liu}}, \bibinfo {author} {\bibfnamefont {Y.}~\bibnamefont {Xia}}, \ and\
  \bibinfo {author} {\bibfnamefont {L.}~\bibnamefont {Mei}},\ }\bibfield
  {title} {\enquote {\bibinfo {title} {Concentration dependent magnetism
  induced by hydrogen adsorption on graphene and single walled carbon
  nanotubes},}\ }\href {\doibase 10.1016/j.jmmm.2009.11.014} {\bibfield
  {journal} {\bibinfo  {journal} {J. Magn. Magn. Mater.}\ }\textbf {\bibinfo
  {volume} {322}},\ \bibinfo {pages} {838--843} (\bibinfo {year}
  {2010})}\BibitemShut {NoStop}%
\bibitem [{\citenamefont {Boukhvalov}, \citenamefont {Katsnelson},\ and\
  \citenamefont {Lichtenstein}(2008)}]{PhysRevB.77.035427}%
  \BibitemOpen
  \bibfield  {author} {\bibinfo {author} {\bibfnamefont {D.~W.}\ \bibnamefont
  {Boukhvalov}}, \bibinfo {author} {\bibfnamefont {M.~I.}\ \bibnamefont
  {Katsnelson}}, \ and\ \bibinfo {author} {\bibfnamefont {A.~I.}\ \bibnamefont
  {Lichtenstein}},\ }\bibfield  {title} {\enquote {\bibinfo {title} {Hydrogen
  on graphene: Electronic structure, total energy, structural distortions and
  magnetism from first-principles calculations},}\ }\href {\doibase
  10.1103/PhysRevB.77.035427} {\bibfield  {journal} {\bibinfo  {journal} {Phys.
  Rev. B}\ }\textbf {\bibinfo {volume} {77}},\ \bibinfo {pages} {035427}
  (\bibinfo {year} {2008})}\BibitemShut {NoStop}%
\bibitem [{Note1()}]{Note1}%
  \BibitemOpen
  \bibinfo {note} {Under review: M. A. Kim, D. Sorescu, S. Amemiya, K. D.
  Jordan, and H. Liu, ``Real Time Modulation of Hydrogen Evolution Activity of
  Graphene Electrodes Using Mechanical Strain,'' \protect \textit {ACS Appl.
  Mater. Interfaces}.}\BibitemShut {Stop}%
\bibitem [{\citenamefont {Grimme}\ \emph {et~al.}(2010)\citenamefont {Grimme},
  \citenamefont {Antony}, \citenamefont {Ehrlich},\ and\ \citenamefont
  {Krieg}}]{doi:10.1063/1.3382344}%
  \BibitemOpen
  \bibfield  {author} {\bibinfo {author} {\bibfnamefont {S.}~\bibnamefont
  {Grimme}}, \bibinfo {author} {\bibfnamefont {J.}~\bibnamefont {Antony}},
  \bibinfo {author} {\bibfnamefont {S.}~\bibnamefont {Ehrlich}}, \ and\
  \bibinfo {author} {\bibfnamefont {H.}~\bibnamefont {Krieg}},\ }\bibfield
  {title} {\enquote {\bibinfo {title} {{A consistent and accurate ab initio
  parametrization of density functional dispersion correction (DFT-D) for the
  94 elements H-Pu}},}\ }\href {\doibase 10.1063/1.3382344} {\bibfield
  {journal} {\bibinfo  {journal} {J. Chem. Phys.}\ }\textbf {\bibinfo {volume}
  {132}},\ \bibinfo {pages} {154104} (\bibinfo {year} {2010})}\BibitemShut
  {NoStop}%
\bibitem [{\citenamefont {Annaberdiyev}\ \emph {et~al.}(2018)\citenamefont
  {Annaberdiyev}, \citenamefont {Wang}, \citenamefont {Melton}, \citenamefont
  {Bennett}, \citenamefont {Shulenburger},\ and\ \citenamefont
  {Mitas}}]{ccecp_1}%
  \BibitemOpen
  \bibfield  {author} {\bibinfo {author} {\bibfnamefont {A.}~\bibnamefont
  {Annaberdiyev}}, \bibinfo {author} {\bibfnamefont {G.}~\bibnamefont {Wang}},
  \bibinfo {author} {\bibfnamefont {C.~A.}\ \bibnamefont {Melton}}, \bibinfo
  {author} {\bibfnamefont {M.~C.}\ \bibnamefont {Bennett}}, \bibinfo {author}
  {\bibfnamefont {L.}~\bibnamefont {Shulenburger}}, \ and\ \bibinfo {author}
  {\bibfnamefont {L.}~\bibnamefont {Mitas}},\ }\bibfield  {title} {\enquote
  {\bibinfo {title} {{A new generation of effective core potentials from
  correlated calculations: 3d transition metal series}},}\ }\href {\doibase
  10.1063/1.5040472} {\bibfield  {journal} {\bibinfo  {journal} {J. Chem.
  Phys.}\ }\textbf {\bibinfo {volume} {149}},\ \bibinfo {pages} {134108}
  (\bibinfo {year} {2018})}\BibitemShut {NoStop}%
\bibitem [{\citenamefont {Bennett}\ \emph {et~al.}(2017)\citenamefont
  {Bennett}, \citenamefont {Melton}, \citenamefont {Annaberdiyev},
  \citenamefont {Wang}, \citenamefont {Shulenburger},\ and\ \citenamefont
  {Mitas}}]{ccecp_2}%
  \BibitemOpen
  \bibfield  {author} {\bibinfo {author} {\bibfnamefont {M.~C.}\ \bibnamefont
  {Bennett}}, \bibinfo {author} {\bibfnamefont {C.~A.}\ \bibnamefont {Melton}},
  \bibinfo {author} {\bibfnamefont {A.}~\bibnamefont {Annaberdiyev}}, \bibinfo
  {author} {\bibfnamefont {G.}~\bibnamefont {Wang}}, \bibinfo {author}
  {\bibfnamefont {L.}~\bibnamefont {Shulenburger}}, \ and\ \bibinfo {author}
  {\bibfnamefont {L.}~\bibnamefont {Mitas}},\ }\bibfield  {title} {\enquote
  {\bibinfo {title} {{A new generation of effective core potentials for
  correlated calculations}},}\ }\href {\doibase 10.1063/1.4995643} {\bibfield
  {journal} {\bibinfo  {journal} {J. Chem. Phys.}\ }\textbf {\bibinfo {volume}
  {147}},\ \bibinfo {pages} {224106} (\bibinfo {year} {2017})}\BibitemShut
  {NoStop}%
\bibitem [{\citenamefont {Monkhorst}\ and\ \citenamefont
  {Pack}(1976)}]{PhysRevB.13.5188}%
  \BibitemOpen
  \bibfield  {author} {\bibinfo {author} {\bibfnamefont {H.~J.}\ \bibnamefont
  {Monkhorst}}\ and\ \bibinfo {author} {\bibfnamefont {J.~D.}\ \bibnamefont
  {Pack}},\ }\bibfield  {title} {\enquote {\bibinfo {title} {{Special points
  for Brillouin-zone integrations}},}\ }\href {\doibase
  10.1103/PhysRevB.13.5188} {\bibfield  {journal} {\bibinfo  {journal} {Phys.
  Rev. B}\ }\textbf {\bibinfo {volume} {13}},\ \bibinfo {pages} {5188--5192}
  (\bibinfo {year} {1976})}\BibitemShut {NoStop}%
\bibitem [{\citenamefont {Marzari}\ \emph {et~al.}(1999)\citenamefont
  {Marzari}, \citenamefont {Vanderbilt}, \citenamefont {De~Vita},\ and\
  \citenamefont {Payne}}]{PhysRevLett.82.3296}%
  \BibitemOpen
  \bibfield  {author} {\bibinfo {author} {\bibfnamefont {N.}~\bibnamefont
  {Marzari}}, \bibinfo {author} {\bibfnamefont {D.}~\bibnamefont {Vanderbilt}},
  \bibinfo {author} {\bibfnamefont {A.}~\bibnamefont {De~Vita}}, \ and\
  \bibinfo {author} {\bibfnamefont {M.~C.}\ \bibnamefont {Payne}},\ }\bibfield
  {title} {\enquote {\bibinfo {title} {{Thermal Contraction and Disordering of
  the Al(110) Surface}},}\ }\href {\doibase 10.1103/PhysRevLett.82.3296}
  {\bibfield  {journal} {\bibinfo  {journal} {Phys. Rev. Lett.}\ }\textbf
  {\bibinfo {volume} {82}},\ \bibinfo {pages} {3296--3299} (\bibinfo {year}
  {1999})}\BibitemShut {NoStop}%
\bibitem [{\citenamefont {Adamo}\ and\ \citenamefont
  {Barone}(1999)}]{doi:10.1063/1.478522}%
  \BibitemOpen
  \bibfield  {author} {\bibinfo {author} {\bibfnamefont {C.}~\bibnamefont
  {Adamo}}\ and\ \bibinfo {author} {\bibfnamefont {V.}~\bibnamefont {Barone}},\
  }\bibfield  {title} {\enquote {\bibinfo {title} {{Toward reliable density
  functional methods without adjustable parameters: The PBE0 model}},}\ }\href
  {\doibase 10.1063/1.478522} {\bibfield  {journal} {\bibinfo  {journal} {J.
  Chem. Phys.}\ }\textbf {\bibinfo {volume} {110}},\ \bibinfo {pages}
  {6158--6170} (\bibinfo {year} {1999})}\BibitemShut {NoStop}%
\bibitem [{\citenamefont {Krukau}\ \emph {et~al.}(2006)\citenamefont {Krukau},
  \citenamefont {Vydrov}, \citenamefont {Izmaylov},\ and\ \citenamefont
  {Scuseria}}]{doi:10.1063/1.2404663}%
  \BibitemOpen
  \bibfield  {author} {\bibinfo {author} {\bibfnamefont {A.~V.}\ \bibnamefont
  {Krukau}}, \bibinfo {author} {\bibfnamefont {O.~A.}\ \bibnamefont {Vydrov}},
  \bibinfo {author} {\bibfnamefont {A.~F.}\ \bibnamefont {Izmaylov}}, \ and\
  \bibinfo {author} {\bibfnamefont {G.~E.}\ \bibnamefont {Scuseria}},\
  }\bibfield  {title} {\enquote {\bibinfo {title} {Influence of the exchange
  screening parameter on the performance of screened hybrid functionals},}\
  }\href {\doibase 10.1063/1.2404663} {\bibfield  {journal} {\bibinfo
  {journal} {J. Chem. Phys.}\ }\textbf {\bibinfo {volume} {125}},\ \bibinfo
  {pages} {224106} (\bibinfo {year} {2006})}\BibitemShut {NoStop}%
\bibitem [{\citenamefont {Peintinger}, \citenamefont {Oliveira},\ and\
  \citenamefont {Bredow}(2012)}]{Peintinger2012-ff}%
  \BibitemOpen
  \bibfield  {author} {\bibinfo {author} {\bibfnamefont {M.~F.}\ \bibnamefont
  {Peintinger}}, \bibinfo {author} {\bibfnamefont {D.~V.}\ \bibnamefont
  {Oliveira}}, \ and\ \bibinfo {author} {\bibfnamefont {T.}~\bibnamefont
  {Bredow}},\ }\bibfield  {title} {\enquote {\bibinfo {title} {Consistent
  gaussian basis sets of triple-zeta valence with polarization quality for
  solid-state calculations},}\ }\href@noop {} {\bibfield  {journal} {\bibinfo
  {journal} {J. Comput. Chem.}\ }\textbf {\bibinfo {volume} {34}},\ \bibinfo
  {pages} {451--459} (\bibinfo {year} {2012})}\BibitemShut {NoStop}%
\bibitem [{\citenamefont {Kruse}\ and\ \citenamefont
  {Grimme}(2012)}]{KruseGrimme2012}%
  \BibitemOpen
  \bibfield  {author} {\bibinfo {author} {\bibfnamefont {H.}~\bibnamefont
  {Kruse}}\ and\ \bibinfo {author} {\bibfnamefont {S.}~\bibnamefont {Grimme}},\
  }\bibfield  {title} {\enquote {\bibinfo {title} {{A geometrical correction
  for the inter- and intra-molecular basis set superposition error in
  Hartree-Fock and density functional theory calculations for large
  systems}},}\ }\href {\doibase 10.1063/1.3700154} {\bibfield  {journal}
  {\bibinfo  {journal} {J. Chem. Phys.}\ }\textbf {\bibinfo {volume} {136}},\
  \bibinfo {pages} {154101} (\bibinfo {year} {2012})}\BibitemShut {NoStop}%
\bibitem [{\citenamefont {Brandenburg}\ \emph {et~al.}(2013)\citenamefont
  {Brandenburg}, \citenamefont {Alessio}, \citenamefont {Civalleri},
  \citenamefont {Peintinger}, \citenamefont {Bredow},\ and\ \citenamefont
  {Grimme}}]{Brandenburg2013}%
  \BibitemOpen
  \bibfield  {author} {\bibinfo {author} {\bibfnamefont {J.~G.}\ \bibnamefont
  {Brandenburg}}, \bibinfo {author} {\bibfnamefont {M.}~\bibnamefont
  {Alessio}}, \bibinfo {author} {\bibfnamefont {B.}~\bibnamefont {Civalleri}},
  \bibinfo {author} {\bibfnamefont {M.~F.}\ \bibnamefont {Peintinger}},
  \bibinfo {author} {\bibfnamefont {T.}~\bibnamefont {Bredow}}, \ and\ \bibinfo
  {author} {\bibfnamefont {S.}~\bibnamefont {Grimme}},\ }\bibfield  {title}
  {\enquote {\bibinfo {title} {{Geometrical Correction for the Inter- and
  Intramolecular Basis Set Superposition Error in Periodic Density Functional
  Theory Calculations}},}\ }\href {\doibase 10.1021/jp406658y} {\bibfield
  {journal} {\bibinfo  {journal} {J. Phys. Chem. A}\ }\textbf {\bibinfo
  {volume} {117}},\ \bibinfo {pages} {9282--9292} (\bibinfo {year}
  {2013})}\BibitemShut {NoStop}%
\bibitem [{\citenamefont {Anderson}(1980)}]{Anderson1980}%
  \BibitemOpen
  \bibfield  {author} {\bibinfo {author} {\bibfnamefont {J.~B.}\ \bibnamefont
  {Anderson}},\ }\bibfield  {title} {\enquote {\bibinfo {title} {Quantum
  chemistry by random walk: Higher accuracy},}\ }\href {\doibase
  10.1063/1.440575} {\bibfield  {journal} {\bibinfo  {journal} {J. Chem.
  Phys.}\ }\textbf {\bibinfo {volume} {73}},\ \bibinfo {pages} {3897--3899}
  (\bibinfo {year} {1980})}\BibitemShut {NoStop}%
\bibitem [{\citenamefont {Umrigar}\ \emph {et~al.}(2007)\citenamefont
  {Umrigar}, \citenamefont {Toulouse}, \citenamefont {Filippi}, \citenamefont
  {Sorella},\ and\ \citenamefont {Hennig}}]{10.1103/PhysRevLett.98.110201}%
  \BibitemOpen
  \bibfield  {author} {\bibinfo {author} {\bibfnamefont {C.~J.}\ \bibnamefont
  {Umrigar}}, \bibinfo {author} {\bibfnamefont {J.}~\bibnamefont {Toulouse}},
  \bibinfo {author} {\bibfnamefont {C.}~\bibnamefont {Filippi}}, \bibinfo
  {author} {\bibfnamefont {S.}~\bibnamefont {Sorella}}, \ and\ \bibinfo
  {author} {\bibfnamefont {R.~G.}\ \bibnamefont {Hennig}},\ }\bibfield  {title}
  {\enquote {\bibinfo {title} {{Alleviation of the Fermion-Sign Problem by
  Optimization of Many-Body Wave Functions}},}\ }\href {\doibase
  10.1103/PhysRevLett.98.110201} {\bibfield  {journal} {\bibinfo  {journal}
  {Phys. Rev. Lett.}\ }\textbf {\bibinfo {volume} {98}},\ \bibinfo {pages}
  {110201} (\bibinfo {year} {2007})}\BibitemShut {NoStop}%
\bibitem [{\citenamefont {Zen}\ \emph {et~al.}(2019)\citenamefont {Zen},
  \citenamefont {Brandenburg}, \citenamefont {Michaelides},\ and\ \citenamefont
  {Alf\`{e}}}]{zen_new_2019}%
  \BibitemOpen
  \bibfield  {author} {\bibinfo {author} {\bibfnamefont {A.}~\bibnamefont
  {Zen}}, \bibinfo {author} {\bibfnamefont {J.~G.}\ \bibnamefont
  {Brandenburg}}, \bibinfo {author} {\bibfnamefont {A.}~\bibnamefont
  {Michaelides}}, \ and\ \bibinfo {author} {\bibfnamefont {D.}~\bibnamefont
  {Alf\`{e}}},\ }\bibfield  {title} {\enquote {\bibinfo {title} {{A new scheme
  for fixed node diffusion quantum Monte Carlo with pseudopotentials: Improving
  reproducibility and reducing the trial-wave-function bias}},}\ }\href
  {\doibase 10.1063/1.5119729} {\bibfield  {journal} {\bibinfo  {journal} {J.
  Chem. Phys.}\ }\textbf {\bibinfo {volume} {151}},\ \bibinfo {pages} {134105}
  (\bibinfo {year} {2019})}\BibitemShut {NoStop}%
\bibitem [{\citenamefont {Casula}\ \emph {et~al.}(2010)\citenamefont {Casula},
  \citenamefont {Moroni}, \citenamefont {Sorella},\ and\ \citenamefont
  {Filippi}}]{10.1063/1.3380831}%
  \BibitemOpen
  \bibfield  {author} {\bibinfo {author} {\bibfnamefont {M.}~\bibnamefont
  {Casula}}, \bibinfo {author} {\bibfnamefont {S.}~\bibnamefont {Moroni}},
  \bibinfo {author} {\bibfnamefont {S.}~\bibnamefont {Sorella}}, \ and\
  \bibinfo {author} {\bibfnamefont {C.}~\bibnamefont {Filippi}},\ }\bibfield
  {title} {\enquote {\bibinfo {title} {{Size-consistent variational approaches
  to nonlocal pseudopotentials: Standard and lattice regularized diffusion
  Monte Carlo methods revisited}},}\ }\href {\doibase 10.1063/1.3380831}
  {\bibfield  {journal} {\bibinfo  {journal} {J. Chem. Phys.}\ }\textbf
  {\bibinfo {volume} {132}},\ \bibinfo {pages} {154113} (\bibinfo {year}
  {2010})}\BibitemShut {NoStop}%
\bibitem [{\citenamefont {Lin}, \citenamefont {Zong},\ and\ \citenamefont
  {Ceperley}(2001)}]{PhysRevE.64.016702}%
  \BibitemOpen
  \bibfield  {author} {\bibinfo {author} {\bibfnamefont {C.}~\bibnamefont
  {Lin}}, \bibinfo {author} {\bibfnamefont {F.~H.}\ \bibnamefont {Zong}}, \
  and\ \bibinfo {author} {\bibfnamefont {D.~M.}\ \bibnamefont {Ceperley}},\
  }\bibfield  {title} {\enquote {\bibinfo {title} {Twist-averaged boundary
  conditions in continuum quantum monte carlo algorithms},}\ }\href {\doibase
  10.1103/PhysRevE.64.016702} {\bibfield  {journal} {\bibinfo  {journal} {Phys.
  Rev. E}\ }\textbf {\bibinfo {volume} {64}},\ \bibinfo {pages} {016702}
  (\bibinfo {year} {2001})}\BibitemShut {NoStop}%
\bibitem [{\citenamefont {Zen}\ \emph {et~al.}(2016)\citenamefont {Zen},
  \citenamefont {Sorella}, \citenamefont {Gillan}, \citenamefont
  {Michaelides},\ and\ \citenamefont {Alf\`e}}]{10.1103/PhysRevB.93.241118}%
  \BibitemOpen
  \bibfield  {author} {\bibinfo {author} {\bibfnamefont {A.}~\bibnamefont
  {Zen}}, \bibinfo {author} {\bibfnamefont {S.}~\bibnamefont {Sorella}},
  \bibinfo {author} {\bibfnamefont {M.~J.}\ \bibnamefont {Gillan}}, \bibinfo
  {author} {\bibfnamefont {A.}~\bibnamefont {Michaelides}}, \ and\ \bibinfo
  {author} {\bibfnamefont {D.}~\bibnamefont {Alf\`e}},\ }\bibfield  {title}
  {\enquote {\bibinfo {title} {{Boosting the accuracy and speed of quantum
  Monte Carlo: Size consistency and time step}},}\ }\href {\doibase
  10.1103/PhysRevB.93.241118} {\bibfield  {journal} {\bibinfo  {journal} {Phys.
  Rev. B}\ }\textbf {\bibinfo {volume} {93}},\ \bibinfo {pages} {241118}
  (\bibinfo {year} {2016})}\BibitemShut {NoStop}%
\bibitem [{\citenamefont {Shin}\ \emph
  {et~al.}(2019{\natexlab{b}})\citenamefont {Shin}, \citenamefont {Luo},
  \citenamefont {Benali},\ and\ \citenamefont
  {Kwon}}]{10.1103/PhysRevB.100.075430}%
  \BibitemOpen
  \bibfield  {author} {\bibinfo {author} {\bibfnamefont {H.}~\bibnamefont
  {Shin}}, \bibinfo {author} {\bibfnamefont {Y.}~\bibnamefont {Luo}}, \bibinfo
  {author} {\bibfnamefont {A.}~\bibnamefont {Benali}}, \ and\ \bibinfo {author}
  {\bibfnamefont {Y.}~\bibnamefont {Kwon}},\ }\bibfield  {title} {\enquote
  {\bibinfo {title} {{Diffusion Monte Carlo study of ${\mathrm{O}}_{2}$
  adsorption on single layer graphene}},}\ }\href {\doibase
  10.1103/PhysRevB.100.075430} {\bibfield  {journal} {\bibinfo  {journal}
  {Phys. Rev. B}\ }\textbf {\bibinfo {volume} {100}},\ \bibinfo {pages}
  {075430} (\bibinfo {year} {2019}{\natexlab{b}})}\BibitemShut {NoStop}%
\bibitem [{\citenamefont {Cortés-Arriagada}\ \emph {et~al.}(2014)\citenamefont
  {Cortés-Arriagada}, \citenamefont {Gutiérrez-Oliva}, \citenamefont
  {Herrera}, \citenamefont {Soto},\ and\ \citenamefont
  {Toro-Labbé}}]{doi:10.1063/1.4896611}%
  \BibitemOpen
  \bibfield  {author} {\bibinfo {author} {\bibfnamefont {D.}~\bibnamefont
  {Cortés-Arriagada}}, \bibinfo {author} {\bibfnamefont {S.}~\bibnamefont
  {Gutiérrez-Oliva}}, \bibinfo {author} {\bibfnamefont {B.}~\bibnamefont
  {Herrera}}, \bibinfo {author} {\bibfnamefont {K.}~\bibnamefont {Soto}}, \
  and\ \bibinfo {author} {\bibfnamefont {A.}~\bibnamefont {Toro-Labbé}},\
  }\bibfield  {title} {\enquote {\bibinfo {title} {{The mechanism of
  chemisorption of hydrogen atom on graphene: Insights from the reaction force
  and reaction electronic flux}},}\ }\href {\doibase 10.1063/1.4896611}
  {\bibfield  {journal} {\bibinfo  {journal} {J. Chem. Phys.}\ }\textbf
  {\bibinfo {volume} {141}},\ \bibinfo {pages} {134701} (\bibinfo {year}
  {2014})}\BibitemShut {NoStop}%
\bibitem [{\citenamefont {Jiang}\ \emph {et~al.}(2019)\citenamefont {Jiang},
  \citenamefont {Kammler}, \citenamefont {Ding}, \citenamefont {Dorenkamp},
  \citenamefont {Manby}, \citenamefont {Wodtke}, \citenamefont {Miller},
  \citenamefont {Kandratsenka},\ and\ \citenamefont
  {Bünermann}}]{doi:10.1126/science.aaw6378}%
  \BibitemOpen
  \bibfield  {author} {\bibinfo {author} {\bibfnamefont {H.}~\bibnamefont
  {Jiang}}, \bibinfo {author} {\bibfnamefont {M.}~\bibnamefont {Kammler}},
  \bibinfo {author} {\bibfnamefont {F.}~\bibnamefont {Ding}}, \bibinfo {author}
  {\bibfnamefont {Y.}~\bibnamefont {Dorenkamp}}, \bibinfo {author}
  {\bibfnamefont {F.~R.}\ \bibnamefont {Manby}}, \bibinfo {author}
  {\bibfnamefont {A.~M.}\ \bibnamefont {Wodtke}}, \bibinfo {author}
  {\bibfnamefont {T.~F.}\ \bibnamefont {Miller}}, \bibinfo {author}
  {\bibfnamefont {A.}~\bibnamefont {Kandratsenka}}, \ and\ \bibinfo {author}
  {\bibfnamefont {O.}~\bibnamefont {Bünermann}},\ }\bibfield  {title}
  {\enquote {\bibinfo {title} {{Imaging covalent bond formation by H atom
  scattering from graphene}},}\ }\href {\doibase 10.1126/science.aaw6378}
  {\bibfield  {journal} {\bibinfo  {journal} {Science}\ }\textbf {\bibinfo
  {volume} {364}},\ \bibinfo {pages} {379--382} (\bibinfo {year}
  {2019})}\BibitemShut {NoStop}%
\bibitem [{\citenamefont {Giannozzi}\ \emph {et~al.}(2020)\citenamefont
  {Giannozzi}, \citenamefont {Baseggio}, \citenamefont {Bonfà}, \citenamefont
  {Brunato}, \citenamefont {Car}, \citenamefont {Carnimeo}, \citenamefont
  {Cavazzoni}, \citenamefont {de~Gironcoli}, \citenamefont {Delugas},
  \citenamefont {Ferrari~Ruffino}, \citenamefont {Ferretti}, \citenamefont
  {Marzari}, \citenamefont {Timrov}, \citenamefont {Urru},\ and\ \citenamefont
  {Baroni}}]{doi:10.1063/5.0005082}%
  \BibitemOpen
  \bibfield  {author} {\bibinfo {author} {\bibfnamefont {P.}~\bibnamefont
  {Giannozzi}}, \bibinfo {author} {\bibfnamefont {O.}~\bibnamefont {Baseggio}},
  \bibinfo {author} {\bibfnamefont {P.}~\bibnamefont {Bonfà}}, \bibinfo
  {author} {\bibfnamefont {D.}~\bibnamefont {Brunato}}, \bibinfo {author}
  {\bibfnamefont {R.}~\bibnamefont {Car}}, \bibinfo {author} {\bibfnamefont
  {I.}~\bibnamefont {Carnimeo}}, \bibinfo {author} {\bibfnamefont
  {C.}~\bibnamefont {Cavazzoni}}, \bibinfo {author} {\bibfnamefont
  {S.}~\bibnamefont {de~Gironcoli}}, \bibinfo {author} {\bibfnamefont
  {P.}~\bibnamefont {Delugas}}, \bibinfo {author} {\bibfnamefont
  {F.}~\bibnamefont {Ferrari~Ruffino}}, \bibinfo {author} {\bibfnamefont
  {A.}~\bibnamefont {Ferretti}}, \bibinfo {author} {\bibfnamefont
  {N.}~\bibnamefont {Marzari}}, \bibinfo {author} {\bibfnamefont
  {I.}~\bibnamefont {Timrov}}, \bibinfo {author} {\bibfnamefont
  {A.}~\bibnamefont {Urru}}, \ and\ \bibinfo {author} {\bibfnamefont
  {S.}~\bibnamefont {Baroni}},\ }\bibfield  {title} {\enquote {\bibinfo {title}
  {{Quantum ESPRESSO toward the exascale}},}\ }\href {\doibase
  10.1063/5.0005082} {\bibfield  {journal} {\bibinfo  {journal} {J. Chem.
  Phys.}\ }\textbf {\bibinfo {volume} {152}},\ \bibinfo {pages} {154105}
  (\bibinfo {year} {2020})}\BibitemShut {NoStop}%
\bibitem [{\citenamefont {Giannozzi}\ \emph {et~al.}(2009)\citenamefont
  {Giannozzi}, \citenamefont {Baroni}, \citenamefont {Bonini}, \citenamefont
  {Calandra}, \citenamefont {Car}, \citenamefont {Cavazzoni}, \citenamefont
  {Ceresoli}, \citenamefont {Chiarotti}, \citenamefont {Cococcioni},
  \citenamefont {Dabo}, \citenamefont {{Dal Corso}}, \citenamefont
  {de~Gironcoli}, \citenamefont {Fabris}, \citenamefont {Fratesi},
  \citenamefont {Gebauer}, \citenamefont {Gerstmann}, \citenamefont
  {Gougoussis}, \citenamefont {Kokalj}, \citenamefont {Lazzeri}, \citenamefont
  {Martin-Samos}, \citenamefont {Marzari}, \citenamefont {Mauri}, \citenamefont
  {Mazzarello}, \citenamefont {Paolini}, \citenamefont {Pasquarello},
  \citenamefont {Paulatto}, \citenamefont {Sbraccia}, \citenamefont {Scandolo},
  \citenamefont {Sclauzero}, \citenamefont {Seitsonen}, \citenamefont
  {Smogunov}, \citenamefont {Umari},\ and\ \citenamefont
  {Wentzcovitch}}]{QE-2009}%
  \BibitemOpen
  \bibfield  {author} {\bibinfo {author} {\bibfnamefont {P.}~\bibnamefont
  {Giannozzi}}, \bibinfo {author} {\bibfnamefont {S.}~\bibnamefont {Baroni}},
  \bibinfo {author} {\bibfnamefont {N.}~\bibnamefont {Bonini}}, \bibinfo
  {author} {\bibfnamefont {M.}~\bibnamefont {Calandra}}, \bibinfo {author}
  {\bibfnamefont {R.}~\bibnamefont {Car}}, \bibinfo {author} {\bibfnamefont
  {C.}~\bibnamefont {Cavazzoni}}, \bibinfo {author} {\bibfnamefont
  {D.}~\bibnamefont {Ceresoli}}, \bibinfo {author} {\bibfnamefont {G.~L.}\
  \bibnamefont {Chiarotti}}, \bibinfo {author} {\bibfnamefont {M.}~\bibnamefont
  {Cococcioni}}, \bibinfo {author} {\bibfnamefont {I.}~\bibnamefont {Dabo}},
  \bibinfo {author} {\bibfnamefont {A.}~\bibnamefont {{Dal Corso}}}, \bibinfo
  {author} {\bibfnamefont {S.}~\bibnamefont {de~Gironcoli}}, \bibinfo {author}
  {\bibfnamefont {S.}~\bibnamefont {Fabris}}, \bibinfo {author} {\bibfnamefont
  {G.}~\bibnamefont {Fratesi}}, \bibinfo {author} {\bibfnamefont
  {R.}~\bibnamefont {Gebauer}}, \bibinfo {author} {\bibfnamefont
  {U.}~\bibnamefont {Gerstmann}}, \bibinfo {author} {\bibfnamefont
  {C.}~\bibnamefont {Gougoussis}}, \bibinfo {author} {\bibfnamefont
  {A.}~\bibnamefont {Kokalj}}, \bibinfo {author} {\bibfnamefont
  {M.}~\bibnamefont {Lazzeri}}, \bibinfo {author} {\bibfnamefont
  {L.}~\bibnamefont {Martin-Samos}}, \bibinfo {author} {\bibfnamefont
  {N.}~\bibnamefont {Marzari}}, \bibinfo {author} {\bibfnamefont
  {F.}~\bibnamefont {Mauri}}, \bibinfo {author} {\bibfnamefont
  {R.}~\bibnamefont {Mazzarello}}, \bibinfo {author} {\bibfnamefont
  {S.}~\bibnamefont {Paolini}}, \bibinfo {author} {\bibfnamefont
  {A.}~\bibnamefont {Pasquarello}}, \bibinfo {author} {\bibfnamefont
  {L.}~\bibnamefont {Paulatto}}, \bibinfo {author} {\bibfnamefont
  {C.}~\bibnamefont {Sbraccia}}, \bibinfo {author} {\bibfnamefont
  {S.}~\bibnamefont {Scandolo}}, \bibinfo {author} {\bibfnamefont
  {G.}~\bibnamefont {Sclauzero}}, \bibinfo {author} {\bibfnamefont {A.~P.}\
  \bibnamefont {Seitsonen}}, \bibinfo {author} {\bibfnamefont {A.}~\bibnamefont
  {Smogunov}}, \bibinfo {author} {\bibfnamefont {P.}~\bibnamefont {Umari}}, \
  and\ \bibinfo {author} {\bibfnamefont {R.~M.}\ \bibnamefont {Wentzcovitch}},\
  }\bibfield  {title} {\enquote {\bibinfo {title} {{QUANTUM ESPRESSO: a modular
  and open-source software project for quantum simulations of materials}},}\
  }\href@noop {} {\bibfield  {journal} {\bibinfo  {journal} {{J. Phys.:
  Condens. Matter}}\ }\textbf {\bibinfo {volume} {21}},\ \bibinfo {pages}
  {395502 (19pp)} (\bibinfo {year} {2009})}\BibitemShut {NoStop}%
\bibitem [{\citenamefont {Giannozzi}\ \emph {et~al.}(2017)\citenamefont
  {Giannozzi}, \citenamefont {Andreussi}, \citenamefont {Brumme}, \citenamefont
  {Bunau}, \citenamefont {Nardelli}, \citenamefont {Calandra}, \citenamefont
  {Car}, \citenamefont {Cavazzoni}, \citenamefont {Ceresoli}, \citenamefont
  {Cococcioni}, \citenamefont {Colonna}, \citenamefont {Carnimeo},
  \citenamefont {Corso}, \citenamefont {de~Gironcoli}, \citenamefont {Delugas},
  \citenamefont {Jr}, \citenamefont {Ferretti}, \citenamefont {Floris},
  \citenamefont {Fratesi}, \citenamefont {Fugallo}, \citenamefont {Gebauer},
  \citenamefont {Gerstmann}, \citenamefont {Giustino}, \citenamefont {Gorni},
  \citenamefont {Jia}, \citenamefont {Kawamura}, \citenamefont {Ko},
  \citenamefont {Kokalj}, \citenamefont {Küçükbenli}, \citenamefont
  {Lazzeri}, \citenamefont {Marsili}, \citenamefont {Marzari}, \citenamefont
  {Mauri}, \citenamefont {Nguyen}, \citenamefont {Nguyen}, \citenamefont {de-la
  Roza}, \citenamefont {Paulatto}, \citenamefont {Poncé}, \citenamefont
  {Rocca}, \citenamefont {Sabatini}, \citenamefont {Santra}, \citenamefont
  {Schlipf}, \citenamefont {Seitsonen}, \citenamefont {Smogunov}, \citenamefont
  {Timrov}, \citenamefont {Thonhauser}, \citenamefont {Umari}, \citenamefont
  {Vast}, \citenamefont {Wu},\ and\ \citenamefont {Baroni}}]{QE-2017}%
  \BibitemOpen
  \bibfield  {author} {\bibinfo {author} {\bibfnamefont {P.}~\bibnamefont
  {Giannozzi}}, \bibinfo {author} {\bibfnamefont {O.}~\bibnamefont
  {Andreussi}}, \bibinfo {author} {\bibfnamefont {T.}~\bibnamefont {Brumme}},
  \bibinfo {author} {\bibfnamefont {O.}~\bibnamefont {Bunau}}, \bibinfo
  {author} {\bibfnamefont {M.~B.}\ \bibnamefont {Nardelli}}, \bibinfo {author}
  {\bibfnamefont {M.}~\bibnamefont {Calandra}}, \bibinfo {author}
  {\bibfnamefont {R.}~\bibnamefont {Car}}, \bibinfo {author} {\bibfnamefont
  {C.}~\bibnamefont {Cavazzoni}}, \bibinfo {author} {\bibfnamefont
  {D.}~\bibnamefont {Ceresoli}}, \bibinfo {author} {\bibfnamefont
  {M.}~\bibnamefont {Cococcioni}}, \bibinfo {author} {\bibfnamefont
  {N.}~\bibnamefont {Colonna}}, \bibinfo {author} {\bibfnamefont
  {I.}~\bibnamefont {Carnimeo}}, \bibinfo {author} {\bibfnamefont {A.~D.}\
  \bibnamefont {Corso}}, \bibinfo {author} {\bibfnamefont {S.}~\bibnamefont
  {de~Gironcoli}}, \bibinfo {author} {\bibfnamefont {P.}~\bibnamefont
  {Delugas}}, \bibinfo {author} {\bibfnamefont {R.~A.~D.}\ \bibnamefont {Jr}},
  \bibinfo {author} {\bibfnamefont {A.}~\bibnamefont {Ferretti}}, \bibinfo
  {author} {\bibfnamefont {A.}~\bibnamefont {Floris}}, \bibinfo {author}
  {\bibfnamefont {G.}~\bibnamefont {Fratesi}}, \bibinfo {author} {\bibfnamefont
  {G.}~\bibnamefont {Fugallo}}, \bibinfo {author} {\bibfnamefont
  {R.}~\bibnamefont {Gebauer}}, \bibinfo {author} {\bibfnamefont
  {U.}~\bibnamefont {Gerstmann}}, \bibinfo {author} {\bibfnamefont
  {F.}~\bibnamefont {Giustino}}, \bibinfo {author} {\bibfnamefont
  {T.}~\bibnamefont {Gorni}}, \bibinfo {author} {\bibfnamefont
  {J.}~\bibnamefont {Jia}}, \bibinfo {author} {\bibfnamefont {M.}~\bibnamefont
  {Kawamura}}, \bibinfo {author} {\bibfnamefont {H.-Y.}\ \bibnamefont {Ko}},
  \bibinfo {author} {\bibfnamefont {A.}~\bibnamefont {Kokalj}}, \bibinfo
  {author} {\bibfnamefont {E.}~\bibnamefont {Küçükbenli}}, \bibinfo {author}
  {\bibfnamefont {M.}~\bibnamefont {Lazzeri}}, \bibinfo {author} {\bibfnamefont
  {M.}~\bibnamefont {Marsili}}, \bibinfo {author} {\bibfnamefont
  {N.}~\bibnamefont {Marzari}}, \bibinfo {author} {\bibfnamefont
  {F.}~\bibnamefont {Mauri}}, \bibinfo {author} {\bibfnamefont {N.~L.}\
  \bibnamefont {Nguyen}}, \bibinfo {author} {\bibfnamefont {H.-V.}\
  \bibnamefont {Nguyen}}, \bibinfo {author} {\bibfnamefont {A.~O.}\
  \bibnamefont {de-la Roza}}, \bibinfo {author} {\bibfnamefont
  {L.}~\bibnamefont {Paulatto}}, \bibinfo {author} {\bibfnamefont
  {S.}~\bibnamefont {Poncé}}, \bibinfo {author} {\bibfnamefont
  {D.}~\bibnamefont {Rocca}}, \bibinfo {author} {\bibfnamefont
  {R.}~\bibnamefont {Sabatini}}, \bibinfo {author} {\bibfnamefont
  {B.}~\bibnamefont {Santra}}, \bibinfo {author} {\bibfnamefont
  {M.}~\bibnamefont {Schlipf}}, \bibinfo {author} {\bibfnamefont {A.~P.}\
  \bibnamefont {Seitsonen}}, \bibinfo {author} {\bibfnamefont {A.}~\bibnamefont
  {Smogunov}}, \bibinfo {author} {\bibfnamefont {I.}~\bibnamefont {Timrov}},
  \bibinfo {author} {\bibfnamefont {T.}~\bibnamefont {Thonhauser}}, \bibinfo
  {author} {\bibfnamefont {P.}~\bibnamefont {Umari}}, \bibinfo {author}
  {\bibfnamefont {N.}~\bibnamefont {Vast}}, \bibinfo {author} {\bibfnamefont
  {X.}~\bibnamefont {Wu}}, \ and\ \bibinfo {author} {\bibfnamefont
  {S.}~\bibnamefont {Baroni}},\ }\bibfield  {title} {\enquote {\bibinfo {title}
  {{Advanced capabilities for materials modelling with QUANTUM ESPRESSO}},}\
  }\href {\doibase 10.1088/1361-648X/aa8f79} {\bibfield  {journal} {\bibinfo
  {journal} {J. Phys.: Condens. Matter}\ }\textbf {\bibinfo {volume} {29}},\
  \bibinfo {pages} {465901} (\bibinfo {year} {2017})}\BibitemShut {NoStop}%
\bibitem [{\citenamefont {Dovesi}\ \emph {et~al.}(2018)\citenamefont {Dovesi},
  \citenamefont {Erba}, \citenamefont {Orlando}, \citenamefont
  {Zicovich-Wilson}, \citenamefont {Civalleri}, \citenamefont {Maschio},
  \citenamefont {Rérat}, \citenamefont {Casassa}, \citenamefont {Baima},
  \citenamefont {Salustro},\ and\ \citenamefont {Kirtman}}]{10.1002/wcms.1360}%
  \BibitemOpen
  \bibfield  {author} {\bibinfo {author} {\bibfnamefont {R.}~\bibnamefont
  {Dovesi}}, \bibinfo {author} {\bibfnamefont {A.}~\bibnamefont {Erba}},
  \bibinfo {author} {\bibfnamefont {R.}~\bibnamefont {Orlando}}, \bibinfo
  {author} {\bibfnamefont {C.~M.}\ \bibnamefont {Zicovich-Wilson}}, \bibinfo
  {author} {\bibfnamefont {B.}~\bibnamefont {Civalleri}}, \bibinfo {author}
  {\bibfnamefont {L.}~\bibnamefont {Maschio}}, \bibinfo {author} {\bibfnamefont
  {M.}~\bibnamefont {Rérat}}, \bibinfo {author} {\bibfnamefont
  {S.}~\bibnamefont {Casassa}}, \bibinfo {author} {\bibfnamefont
  {J.}~\bibnamefont {Baima}}, \bibinfo {author} {\bibfnamefont
  {S.}~\bibnamefont {Salustro}}, \ and\ \bibinfo {author} {\bibfnamefont
  {B.}~\bibnamefont {Kirtman}},\ }\bibfield  {title} {\enquote {\bibinfo
  {title} {{Quantum-mechanical condensed matter simulations with CRYSTAL}},}\
  }\href {\doibase 10.1002/wcms.1360} {\bibfield  {journal} {\bibinfo
  {journal} {Wiley Interdiscip. Rev. Comput. Mol. Sci}\ }\textbf {\bibinfo
  {volume} {8}},\ \bibinfo {pages} {e1360} (\bibinfo {year}
  {2018})}\BibitemShut {NoStop}%
\bibitem [{\citenamefont {Dovesi}\ \emph {et~al.}(2020)\citenamefont {Dovesi},
  \citenamefont {Pascale}, \citenamefont {Civalleri}, \citenamefont {Doll},
  \citenamefont {Harrison}, \citenamefont {Bush}, \citenamefont {D’Arco},
  \citenamefont {Noël}, \citenamefont {Rérat}, \citenamefont {Carbonnière},
  \citenamefont {Causà}, \citenamefont {Salustro}, \citenamefont {Lacivita},
  \citenamefont {Kirtman}, \citenamefont {Ferrari}, \citenamefont {Gentile},
  \citenamefont {Baima}, \citenamefont {Ferrero}, \citenamefont {Demichelis},\
  and\ \citenamefont {De~La~Pierre}}]{doi:10.1063/5.0004892}%
  \BibitemOpen
  \bibfield  {author} {\bibinfo {author} {\bibfnamefont {R.}~\bibnamefont
  {Dovesi}}, \bibinfo {author} {\bibfnamefont {F.}~\bibnamefont {Pascale}},
  \bibinfo {author} {\bibfnamefont {B.}~\bibnamefont {Civalleri}}, \bibinfo
  {author} {\bibfnamefont {K.}~\bibnamefont {Doll}}, \bibinfo {author}
  {\bibfnamefont {N.~M.}\ \bibnamefont {Harrison}}, \bibinfo {author}
  {\bibfnamefont {I.}~\bibnamefont {Bush}}, \bibinfo {author} {\bibfnamefont
  {P.}~\bibnamefont {D’Arco}}, \bibinfo {author} {\bibfnamefont
  {Y.}~\bibnamefont {Noël}}, \bibinfo {author} {\bibfnamefont
  {M.}~\bibnamefont {Rérat}}, \bibinfo {author} {\bibfnamefont
  {P.}~\bibnamefont {Carbonnière}}, \bibinfo {author} {\bibfnamefont
  {M.}~\bibnamefont {Causà}}, \bibinfo {author} {\bibfnamefont
  {S.}~\bibnamefont {Salustro}}, \bibinfo {author} {\bibfnamefont
  {V.}~\bibnamefont {Lacivita}}, \bibinfo {author} {\bibfnamefont
  {B.}~\bibnamefont {Kirtman}}, \bibinfo {author} {\bibfnamefont {A.~M.}\
  \bibnamefont {Ferrari}}, \bibinfo {author} {\bibfnamefont {F.~S.}\
  \bibnamefont {Gentile}}, \bibinfo {author} {\bibfnamefont {J.}~\bibnamefont
  {Baima}}, \bibinfo {author} {\bibfnamefont {M.}~\bibnamefont {Ferrero}},
  \bibinfo {author} {\bibfnamefont {R.}~\bibnamefont {Demichelis}}, \ and\
  \bibinfo {author} {\bibfnamefont {M.}~\bibnamefont {De~La~Pierre}},\
  }\bibfield  {title} {\enquote {\bibinfo {title} {{The CRYSTAL code,
  1976–2020 and beyond, a long story}},}\ }\href {\doibase 10.1063/5.0004892}
  {\bibfield  {journal} {\bibinfo  {journal} {J. Chem. Phys.}\ }\textbf
  {\bibinfo {volume} {152}},\ \bibinfo {pages} {204111} (\bibinfo {year}
  {2020})}\BibitemShut {NoStop}%
\bibitem [{\citenamefont {Apr\`{a}}\ \emph {et~al.}(2020)\citenamefont
  {Apr\`{a}}, \citenamefont {Bylaska}, \citenamefont {de~Jong}, \citenamefont
  {Govind}, \citenamefont {Kowalski}, \citenamefont {Straatsma}, \citenamefont
  {Valiev}, \citenamefont {van Dam}, \citenamefont {Alexeev}, \citenamefont
  {Anchell}, \citenamefont {Anisimov}, \citenamefont {Aquino}, \citenamefont
  {Atta-Fynn}, \citenamefont {Autschbach}, \citenamefont {Bauman},
  \citenamefont {Becca}, \citenamefont {Bernholdt}, \citenamefont
  {Bhaskaran-Nair}, \citenamefont {Bogatko}, \citenamefont {Borowski},
  \citenamefont {Boschen}, \citenamefont {Brabec}, \citenamefont {Bruner},
  \citenamefont {Cau\"{e}t}, \citenamefont {Chen}, \citenamefont {Chuev},
  \citenamefont {Cramer}, \citenamefont {Daily}, \citenamefont {Deegan},
  \citenamefont {Dunning}, \citenamefont {Dupuis}, \citenamefont {Dyall},
  \citenamefont {Fann}, \citenamefont {Fischer}, \citenamefont {Fonari},
  \citenamefont {Fr\:{u}chtl}, \citenamefont {Gagliardi}, \citenamefont
  {Garza}, \citenamefont {Gawande}, \citenamefont {Ghosh}, \citenamefont
  {Glaesemann}, \citenamefont {G\"{o}tz}, \citenamefont {Hammond},
  \citenamefont {Helms}, \citenamefont {Hermes}, \citenamefont {Hirao},
  \citenamefont {Hirata}, \citenamefont {Jacquelin}, \citenamefont {Jensen},
  \citenamefont {Johnson}, \citenamefont {Jónsson}, \citenamefont {Kendall},
  \citenamefont {Klemm}, \citenamefont {Kobayashi}, \citenamefont {Konkov},
  \citenamefont {Krishnamoorthy}, \citenamefont {Krishnan}, \citenamefont
  {Lin}, \citenamefont {Lins}, \citenamefont {Littlefield}, \citenamefont
  {Logsdail}, \citenamefont {Lopata}, \citenamefont {Ma}, \citenamefont
  {Marenich}, \citenamefont {Martin~del Campo}, \citenamefont
  {Mejia-Rodriguez}, \citenamefont {Moore}, \citenamefont {Mullin},
  \citenamefont {Nakajima}, \citenamefont {Nascimento}, \citenamefont
  {Nichols}, \citenamefont {Nichols}, \citenamefont {Nieplocha}, \citenamefont
  {Otero-de-la Roza}, \citenamefont {Palmer}, \citenamefont {Panyala},
  \citenamefont {Pirojsirikul}, \citenamefont {Peng}, \citenamefont {Peverati},
  \citenamefont {Pittner}, \citenamefont {Pollack}, \citenamefont {Richard},
  \citenamefont {Sadayappan}, \citenamefont {Schatz}, \citenamefont {Shelton},
  \citenamefont {Silverstein}, \citenamefont {Smith}, \citenamefont {Soares},
  \citenamefont {Song}, \citenamefont {Swart}, \citenamefont {Taylor},
  \citenamefont {Thomas}, \citenamefont {Tipparaju}, \citenamefont {Truhlar},
  \citenamefont {Tsemekhman}, \citenamefont {Van~Voorhis}, \citenamefont
  {V\'{a}zquez-Mayagoitia}, \citenamefont {Verma}, \citenamefont {Villa},
  \citenamefont {Vishnu}, \citenamefont {Vogiatzis}, \citenamefont {Wang},
  \citenamefont {Weare}, \citenamefont {Williamson}, \citenamefont {Windus},
  \citenamefont {Woli\'{n}ski}, \citenamefont {Wong}, \citenamefont {Wu},
  \citenamefont {Yang}, \citenamefont {Yu}, \citenamefont {Zacharias},
  \citenamefont {Zhang}, \citenamefont {Zhao},\ and\ \citenamefont
  {Harrison}}]{NWCHEM}%
  \BibitemOpen
  \bibfield  {author} {\bibinfo {author} {\bibfnamefont {E.}~\bibnamefont
  {Apr\`{a}}}, \bibinfo {author} {\bibfnamefont {E.~J.}\ \bibnamefont
  {Bylaska}}, \bibinfo {author} {\bibfnamefont {W.~A.}\ \bibnamefont
  {de~Jong}}, \bibinfo {author} {\bibfnamefont {N.}~\bibnamefont {Govind}},
  \bibinfo {author} {\bibfnamefont {K.}~\bibnamefont {Kowalski}}, \bibinfo
  {author} {\bibfnamefont {T.~P.}\ \bibnamefont {Straatsma}}, \bibinfo {author}
  {\bibfnamefont {M.}~\bibnamefont {Valiev}}, \bibinfo {author} {\bibfnamefont
  {H.~J.~J.}\ \bibnamefont {van Dam}}, \bibinfo {author} {\bibfnamefont
  {Y.}~\bibnamefont {Alexeev}}, \bibinfo {author} {\bibfnamefont
  {J.}~\bibnamefont {Anchell}}, \bibinfo {author} {\bibfnamefont
  {V.}~\bibnamefont {Anisimov}}, \bibinfo {author} {\bibfnamefont {F.~W.}\
  \bibnamefont {Aquino}}, \bibinfo {author} {\bibfnamefont {R.}~\bibnamefont
  {Atta-Fynn}}, \bibinfo {author} {\bibfnamefont {J.}~\bibnamefont
  {Autschbach}}, \bibinfo {author} {\bibfnamefont {N.~P.}\ \bibnamefont
  {Bauman}}, \bibinfo {author} {\bibfnamefont {J.~C.}\ \bibnamefont {Becca}},
  \bibinfo {author} {\bibfnamefont {D.~E.}\ \bibnamefont {Bernholdt}}, \bibinfo
  {author} {\bibfnamefont {K.}~\bibnamefont {Bhaskaran-Nair}}, \bibinfo
  {author} {\bibfnamefont {S.}~\bibnamefont {Bogatko}}, \bibinfo {author}
  {\bibfnamefont {P.}~\bibnamefont {Borowski}}, \bibinfo {author}
  {\bibfnamefont {J.}~\bibnamefont {Boschen}}, \bibinfo {author} {\bibfnamefont
  {J.}~\bibnamefont {Brabec}}, \bibinfo {author} {\bibfnamefont
  {A.}~\bibnamefont {Bruner}}, \bibinfo {author} {\bibfnamefont
  {E.}~\bibnamefont {Cau\"{e}t}}, \bibinfo {author} {\bibfnamefont
  {Y.}~\bibnamefont {Chen}}, \bibinfo {author} {\bibfnamefont {G.~N.}\
  \bibnamefont {Chuev}}, \bibinfo {author} {\bibfnamefont {C.~J.}\ \bibnamefont
  {Cramer}}, \bibinfo {author} {\bibfnamefont {J.}~\bibnamefont {Daily}},
  \bibinfo {author} {\bibfnamefont {M.~J.~O.}\ \bibnamefont {Deegan}}, \bibinfo
  {author} {\bibfnamefont {T.~H.}\ \bibnamefont {Dunning}}, \bibinfo {author}
  {\bibfnamefont {M.}~\bibnamefont {Dupuis}}, \bibinfo {author} {\bibfnamefont
  {K.~G.}\ \bibnamefont {Dyall}}, \bibinfo {author} {\bibfnamefont {G.~I.}\
  \bibnamefont {Fann}}, \bibinfo {author} {\bibfnamefont {S.~A.}\ \bibnamefont
  {Fischer}}, \bibinfo {author} {\bibfnamefont {A.}~\bibnamefont {Fonari}},
  \bibinfo {author} {\bibfnamefont {H.}~\bibnamefont {Fr\:{u}chtl}}, \bibinfo
  {author} {\bibfnamefont {L.}~\bibnamefont {Gagliardi}}, \bibinfo {author}
  {\bibfnamefont {J.}~\bibnamefont {Garza}}, \bibinfo {author} {\bibfnamefont
  {N.}~\bibnamefont {Gawande}}, \bibinfo {author} {\bibfnamefont
  {S.}~\bibnamefont {Ghosh}}, \bibinfo {author} {\bibfnamefont
  {K.}~\bibnamefont {Glaesemann}}, \bibinfo {author} {\bibfnamefont {A.~W.}\
  \bibnamefont {G\"{o}tz}}, \bibinfo {author} {\bibfnamefont {J.}~\bibnamefont
  {Hammond}}, \bibinfo {author} {\bibfnamefont {V.}~\bibnamefont {Helms}},
  \bibinfo {author} {\bibfnamefont {E.~D.}\ \bibnamefont {Hermes}}, \bibinfo
  {author} {\bibfnamefont {K.}~\bibnamefont {Hirao}}, \bibinfo {author}
  {\bibfnamefont {S.}~\bibnamefont {Hirata}}, \bibinfo {author} {\bibfnamefont
  {M.}~\bibnamefont {Jacquelin}}, \bibinfo {author} {\bibfnamefont
  {L.}~\bibnamefont {Jensen}}, \bibinfo {author} {\bibfnamefont {B.~G.}\
  \bibnamefont {Johnson}}, \bibinfo {author} {\bibfnamefont {H.}~\bibnamefont
  {Jónsson}}, \bibinfo {author} {\bibfnamefont {R.~A.}\ \bibnamefont
  {Kendall}}, \bibinfo {author} {\bibfnamefont {M.}~\bibnamefont {Klemm}},
  \bibinfo {author} {\bibfnamefont {R.}~\bibnamefont {Kobayashi}}, \bibinfo
  {author} {\bibfnamefont {V.}~\bibnamefont {Konkov}}, \bibinfo {author}
  {\bibfnamefont {S.}~\bibnamefont {Krishnamoorthy}}, \bibinfo {author}
  {\bibfnamefont {M.}~\bibnamefont {Krishnan}}, \bibinfo {author}
  {\bibfnamefont {Z.}~\bibnamefont {Lin}}, \bibinfo {author} {\bibfnamefont
  {R.~D.}\ \bibnamefont {Lins}}, \bibinfo {author} {\bibfnamefont {R.~J.}\
  \bibnamefont {Littlefield}}, \bibinfo {author} {\bibfnamefont {A.~J.}\
  \bibnamefont {Logsdail}}, \bibinfo {author} {\bibfnamefont {K.}~\bibnamefont
  {Lopata}}, \bibinfo {author} {\bibfnamefont {W.}~\bibnamefont {Ma}}, \bibinfo
  {author} {\bibfnamefont {A.~V.}\ \bibnamefont {Marenich}}, \bibinfo {author}
  {\bibfnamefont {J.}~\bibnamefont {Martin~del Campo}}, \bibinfo {author}
  {\bibfnamefont {D.}~\bibnamefont {Mejia-Rodriguez}}, \bibinfo {author}
  {\bibfnamefont {J.~E.}\ \bibnamefont {Moore}}, \bibinfo {author}
  {\bibfnamefont {J.~M.}\ \bibnamefont {Mullin}}, \bibinfo {author}
  {\bibfnamefont {T.}~\bibnamefont {Nakajima}}, \bibinfo {author}
  {\bibfnamefont {D.~R.}\ \bibnamefont {Nascimento}}, \bibinfo {author}
  {\bibfnamefont {J.~A.}\ \bibnamefont {Nichols}}, \bibinfo {author}
  {\bibfnamefont {P.~J.}\ \bibnamefont {Nichols}}, \bibinfo {author}
  {\bibfnamefont {J.}~\bibnamefont {Nieplocha}}, \bibinfo {author}
  {\bibfnamefont {A.}~\bibnamefont {Otero-de-la Roza}}, \bibinfo {author}
  {\bibfnamefont {B.}~\bibnamefont {Palmer}}, \bibinfo {author} {\bibfnamefont
  {A.}~\bibnamefont {Panyala}}, \bibinfo {author} {\bibfnamefont
  {T.}~\bibnamefont {Pirojsirikul}}, \bibinfo {author} {\bibfnamefont
  {B.}~\bibnamefont {Peng}}, \bibinfo {author} {\bibfnamefont {R.}~\bibnamefont
  {Peverati}}, \bibinfo {author} {\bibfnamefont {J.}~\bibnamefont {Pittner}},
  \bibinfo {author} {\bibfnamefont {L.}~\bibnamefont {Pollack}}, \bibinfo
  {author} {\bibfnamefont {R.~M.}\ \bibnamefont {Richard}}, \bibinfo {author}
  {\bibfnamefont {P.}~\bibnamefont {Sadayappan}}, \bibinfo {author}
  {\bibfnamefont {G.~C.}\ \bibnamefont {Schatz}}, \bibinfo {author}
  {\bibfnamefont {W.~A.}\ \bibnamefont {Shelton}}, \bibinfo {author}
  {\bibfnamefont {D.~W.}\ \bibnamefont {Silverstein}}, \bibinfo {author}
  {\bibfnamefont {D.~M.~A.}\ \bibnamefont {Smith}}, \bibinfo {author}
  {\bibfnamefont {T.~A.}\ \bibnamefont {Soares}}, \bibinfo {author}
  {\bibfnamefont {D.}~\bibnamefont {Song}}, \bibinfo {author} {\bibfnamefont
  {M.}~\bibnamefont {Swart}}, \bibinfo {author} {\bibfnamefont {H.~L.}\
  \bibnamefont {Taylor}}, \bibinfo {author} {\bibfnamefont {G.~S.}\
  \bibnamefont {Thomas}}, \bibinfo {author} {\bibfnamefont {V.}~\bibnamefont
  {Tipparaju}}, \bibinfo {author} {\bibfnamefont {D.~G.}\ \bibnamefont
  {Truhlar}}, \bibinfo {author} {\bibfnamefont {K.}~\bibnamefont {Tsemekhman}},
  \bibinfo {author} {\bibfnamefont {T.}~\bibnamefont {Van~Voorhis}}, \bibinfo
  {author} {\bibfnamefont {A.}~\bibnamefont {V\'{a}zquez-Mayagoitia}}, \bibinfo
  {author} {\bibfnamefont {P.}~\bibnamefont {Verma}}, \bibinfo {author}
  {\bibfnamefont {O.}~\bibnamefont {Villa}}, \bibinfo {author} {\bibfnamefont
  {A.}~\bibnamefont {Vishnu}}, \bibinfo {author} {\bibfnamefont {K.~D.}\
  \bibnamefont {Vogiatzis}}, \bibinfo {author} {\bibfnamefont {D.}~\bibnamefont
  {Wang}}, \bibinfo {author} {\bibfnamefont {J.~H.}\ \bibnamefont {Weare}},
  \bibinfo {author} {\bibfnamefont {M.~J.}\ \bibnamefont {Williamson}},
  \bibinfo {author} {\bibfnamefont {T.~L.}\ \bibnamefont {Windus}}, \bibinfo
  {author} {\bibfnamefont {K.}~\bibnamefont {Woli\'{n}ski}}, \bibinfo {author}
  {\bibfnamefont {A.~T.}\ \bibnamefont {Wong}}, \bibinfo {author}
  {\bibfnamefont {Q.}~\bibnamefont {Wu}}, \bibinfo {author} {\bibfnamefont
  {C.}~\bibnamefont {Yang}}, \bibinfo {author} {\bibfnamefont {Q.}~\bibnamefont
  {Yu}}, \bibinfo {author} {\bibfnamefont {M.}~\bibnamefont {Zacharias}},
  \bibinfo {author} {\bibfnamefont {Z.}~\bibnamefont {Zhang}}, \bibinfo
  {author} {\bibfnamefont {Y.}~\bibnamefont {Zhao}}, \ and\ \bibinfo {author}
  {\bibfnamefont {R.~J.}\ \bibnamefont {Harrison}},\ }\bibfield  {title}
  {\enquote {\bibinfo {title} {{NWChem: Past, present, and future}},}\ }\href
  {\doibase 10.1063/5.0004997} {\bibfield  {journal} {\bibinfo  {journal} {J.
  Chem. Phys.}\ }\textbf {\bibinfo {volume} {152}},\ \bibinfo {pages} {184102}
  (\bibinfo {year} {2020})}\BibitemShut {NoStop}%
\bibitem [{\citenamefont {Kim}\ \emph {et~al.}(2018)\citenamefont {Kim},
  \citenamefont {Baczewski}, \citenamefont {Beaudet}, \citenamefont {Benali},
  \citenamefont {Bennett}, \citenamefont {Berrill}, \citenamefont {Blunt},
  \citenamefont {Borda}, \citenamefont {Casula}, \citenamefont {Ceperley},
  \citenamefont {Chiesa}, \citenamefont {Clark}, \citenamefont {Clay},
  \citenamefont {Delaney}, \citenamefont {Dewing}, \citenamefont {Esler},
  \citenamefont {Hao}, \citenamefont {Heinonen}, \citenamefont {Kent},
  \citenamefont {Krogel}, \citenamefont {Kyl{\"{a}}np{\"{a}}{\"{a}}},
  \citenamefont {Li}, \citenamefont {Lopez}, \citenamefont {Luo}, \citenamefont
  {Malone}, \citenamefont {Martin}, \citenamefont {Mathuriya}, \citenamefont
  {McMinis}, \citenamefont {Melton}, \citenamefont {Mitas}, \citenamefont
  {Morales}, \citenamefont {Neuscamman}, \citenamefont {Parker}, \citenamefont
  {Flores}, \citenamefont {Romero}, \citenamefont {Rubenstein}, \citenamefont
  {Shea}, \citenamefont {Shin}, \citenamefont {Shulenburger}, \citenamefont
  {Tillack}, \citenamefont {Townsend}, \citenamefont {Tubman}, \citenamefont
  {Goetz}, \citenamefont {Vincent}, \citenamefont {Yang}, \citenamefont {Yang},
  \citenamefont {Zhang},\ and\ \citenamefont {Zhao}}]{QMCPACK_1}%
  \BibitemOpen
  \bibfield  {author} {\bibinfo {author} {\bibfnamefont {J.}~\bibnamefont
  {Kim}}, \bibinfo {author} {\bibfnamefont {A.~D.}\ \bibnamefont {Baczewski}},
  \bibinfo {author} {\bibfnamefont {T.~D.}\ \bibnamefont {Beaudet}}, \bibinfo
  {author} {\bibfnamefont {A.}~\bibnamefont {Benali}}, \bibinfo {author}
  {\bibfnamefont {M.~C.}\ \bibnamefont {Bennett}}, \bibinfo {author}
  {\bibfnamefont {M.~A.}\ \bibnamefont {Berrill}}, \bibinfo {author}
  {\bibfnamefont {N.~S.}\ \bibnamefont {Blunt}}, \bibinfo {author}
  {\bibfnamefont {E.~J.~L.}\ \bibnamefont {Borda}}, \bibinfo {author}
  {\bibfnamefont {M.}~\bibnamefont {Casula}}, \bibinfo {author} {\bibfnamefont
  {D.~M.}\ \bibnamefont {Ceperley}}, \bibinfo {author} {\bibfnamefont
  {S.}~\bibnamefont {Chiesa}}, \bibinfo {author} {\bibfnamefont {B.~K.}\
  \bibnamefont {Clark}}, \bibinfo {author} {\bibfnamefont {R.~C.}\ \bibnamefont
  {Clay}}, \bibinfo {author} {\bibfnamefont {K.~T.}\ \bibnamefont {Delaney}},
  \bibinfo {author} {\bibfnamefont {M.}~\bibnamefont {Dewing}}, \bibinfo
  {author} {\bibfnamefont {K.~P.}\ \bibnamefont {Esler}}, \bibinfo {author}
  {\bibfnamefont {H.}~\bibnamefont {Hao}}, \bibinfo {author} {\bibfnamefont
  {O.}~\bibnamefont {Heinonen}}, \bibinfo {author} {\bibfnamefont {P.~R.~C.}\
  \bibnamefont {Kent}}, \bibinfo {author} {\bibfnamefont {J.~T.}\ \bibnamefont
  {Krogel}}, \bibinfo {author} {\bibfnamefont {I.}~\bibnamefont
  {Kyl{\"{a}}np{\"{a}}{\"{a}}}}, \bibinfo {author} {\bibfnamefont {Y.~W.}\
  \bibnamefont {Li}}, \bibinfo {author} {\bibfnamefont {M.~G.}\ \bibnamefont
  {Lopez}}, \bibinfo {author} {\bibfnamefont {Y.}~\bibnamefont {Luo}}, \bibinfo
  {author} {\bibfnamefont {F.~D.}\ \bibnamefont {Malone}}, \bibinfo {author}
  {\bibfnamefont {R.~M.}\ \bibnamefont {Martin}}, \bibinfo {author}
  {\bibfnamefont {A.}~\bibnamefont {Mathuriya}}, \bibinfo {author}
  {\bibfnamefont {J.}~\bibnamefont {McMinis}}, \bibinfo {author} {\bibfnamefont
  {C.~A.}\ \bibnamefont {Melton}}, \bibinfo {author} {\bibfnamefont
  {L.}~\bibnamefont {Mitas}}, \bibinfo {author} {\bibfnamefont {M.~A.}\
  \bibnamefont {Morales}}, \bibinfo {author} {\bibfnamefont {E.}~\bibnamefont
  {Neuscamman}}, \bibinfo {author} {\bibfnamefont {W.~D.}\ \bibnamefont
  {Parker}}, \bibinfo {author} {\bibfnamefont {S.~D.~P.}\ \bibnamefont
  {Flores}}, \bibinfo {author} {\bibfnamefont {N.~A.}\ \bibnamefont {Romero}},
  \bibinfo {author} {\bibfnamefont {B.~M.}\ \bibnamefont {Rubenstein}},
  \bibinfo {author} {\bibfnamefont {J.~A.~R.}\ \bibnamefont {Shea}}, \bibinfo
  {author} {\bibfnamefont {H.}~\bibnamefont {Shin}}, \bibinfo {author}
  {\bibfnamefont {L.}~\bibnamefont {Shulenburger}}, \bibinfo {author}
  {\bibfnamefont {A.~F.}\ \bibnamefont {Tillack}}, \bibinfo {author}
  {\bibfnamefont {J.~P.}\ \bibnamefont {Townsend}}, \bibinfo {author}
  {\bibfnamefont {N.~M.}\ \bibnamefont {Tubman}}, \bibinfo {author}
  {\bibfnamefont {B.~V.~D.}\ \bibnamefont {Goetz}}, \bibinfo {author}
  {\bibfnamefont {J.~E.}\ \bibnamefont {Vincent}}, \bibinfo {author}
  {\bibfnamefont {D.~C.}\ \bibnamefont {Yang}}, \bibinfo {author}
  {\bibfnamefont {Y.}~\bibnamefont {Yang}}, \bibinfo {author} {\bibfnamefont
  {S.}~\bibnamefont {Zhang}}, \ and\ \bibinfo {author} {\bibfnamefont
  {L.}~\bibnamefont {Zhao}},\ }\bibfield  {title} {\enquote {\bibinfo {title}
  {{QMCPACK: an open source ab initio quantum Monte Carlo package for the
  electronic structure of atoms, molecules and solids}},}\ }\href {\doibase
  10.1088/1361-648x/aab9c3} {\bibfield  {journal} {\bibinfo  {journal} {J.
  Phys.: Condens. Matter}\ }\textbf {\bibinfo {volume} {30}},\ \bibinfo {pages}
  {195901} (\bibinfo {year} {2018})}\BibitemShut {NoStop}%
\bibitem [{\citenamefont {Kent}\ \emph {et~al.}(2020)\citenamefont {Kent},
  \citenamefont {Annaberdiyev}, \citenamefont {Benali}, \citenamefont
  {Bennett}, \citenamefont {Landinez~Borda}, \citenamefont {Doak},
  \citenamefont {Hao}, \citenamefont {Jordan}, \citenamefont {Krogel},
  \citenamefont {Kyl{{\"{a}}}np{{\"{a}}}{{\"{a}}}}, \citenamefont {Lee},
  \citenamefont {Luo}, \citenamefont {Malone}, \citenamefont {Melton},
  \citenamefont {Mitas}, \citenamefont {Morales}, \citenamefont {Neuscamman},
  \citenamefont {Reboredo}, \citenamefont {Rubenstein}, \citenamefont
  {Saritas}, \citenamefont {Upadhyay}, \citenamefont {Wang}, \citenamefont
  {Zhang},\ and\ \citenamefont {Zhao}}]{QMCPACK_2}%
  \BibitemOpen
  \bibfield  {author} {\bibinfo {author} {\bibfnamefont {P.~R.~C.}\
  \bibnamefont {Kent}}, \bibinfo {author} {\bibfnamefont {A.}~\bibnamefont
  {Annaberdiyev}}, \bibinfo {author} {\bibfnamefont {A.}~\bibnamefont
  {Benali}}, \bibinfo {author} {\bibfnamefont {M.~C.}\ \bibnamefont {Bennett}},
  \bibinfo {author} {\bibfnamefont {E.~J.}\ \bibnamefont {Landinez~Borda}},
  \bibinfo {author} {\bibfnamefont {P.}~\bibnamefont {Doak}}, \bibinfo {author}
  {\bibfnamefont {H.}~\bibnamefont {Hao}}, \bibinfo {author} {\bibfnamefont
  {K.~D.}\ \bibnamefont {Jordan}}, \bibinfo {author} {\bibfnamefont {J.~T.}\
  \bibnamefont {Krogel}}, \bibinfo {author} {\bibfnamefont {I.}~\bibnamefont
  {Kyl{{\"{a}}}np{{\"{a}}}{{\"{a}}}}}, \bibinfo {author} {\bibfnamefont
  {J.}~\bibnamefont {Lee}}, \bibinfo {author} {\bibfnamefont {Y.}~\bibnamefont
  {Luo}}, \bibinfo {author} {\bibfnamefont {F.~D.}\ \bibnamefont {Malone}},
  \bibinfo {author} {\bibfnamefont {C.~A.}\ \bibnamefont {Melton}}, \bibinfo
  {author} {\bibfnamefont {L.}~\bibnamefont {Mitas}}, \bibinfo {author}
  {\bibfnamefont {M.~A.}\ \bibnamefont {Morales}}, \bibinfo {author}
  {\bibfnamefont {E.}~\bibnamefont {Neuscamman}}, \bibinfo {author}
  {\bibfnamefont {F.~A.}\ \bibnamefont {Reboredo}}, \bibinfo {author}
  {\bibfnamefont {B.}~\bibnamefont {Rubenstein}}, \bibinfo {author}
  {\bibfnamefont {K.}~\bibnamefont {Saritas}}, \bibinfo {author} {\bibfnamefont
  {S.}~\bibnamefont {Upadhyay}}, \bibinfo {author} {\bibfnamefont
  {G.}~\bibnamefont {Wang}}, \bibinfo {author} {\bibfnamefont {S.}~\bibnamefont
  {Zhang}}, \ and\ \bibinfo {author} {\bibfnamefont {L.}~\bibnamefont {Zhao}},\
  }\bibfield  {title} {\enquote {\bibinfo {title} {{QMCPACK: Advances in the
  development, efficiency, and application of auxiliary field and real-space
  variational and diffusion quantum Monte Carlo}},}\ }\href {\doibase
  10.1063/5.0004860} {\bibfield  {journal} {\bibinfo  {journal} {J. Chem.
  Phys.}\ }\textbf {\bibinfo {volume} {152}},\ \bibinfo {pages} {174105}
  (\bibinfo {year} {2020})}\BibitemShut {NoStop}%
\bibitem [{\citenamefont {Krogel}(2016)}]{10.1016/j.cpc.2015.08.012}%
  \BibitemOpen
  \bibfield  {author} {\bibinfo {author} {\bibfnamefont {J.~T.}\ \bibnamefont
  {Krogel}},\ }\bibfield  {title} {\enquote {\bibinfo {title} {{Nexus: A
  modular workflow management system for quantum simulation codes}},}\ }\href
  {\doibase 10.1016/j.cpc.2015.08.012} {\bibfield  {journal} {\bibinfo
  {journal} {Comput. Phys. Commun.}\ }\textbf {\bibinfo {volume} {198}},\
  \bibinfo {pages} {154--168} (\bibinfo {year} {2016})}\BibitemShut {NoStop}%
\bibitem [{\citenamefont {Hunter}(2007)}]{matplotlib}%
  \BibitemOpen
  \bibfield  {author} {\bibinfo {author} {\bibfnamefont {J.~D.}\ \bibnamefont
  {Hunter}},\ }\bibfield  {title} {\enquote {\bibinfo {title} {Matplotlib: A 2d
  graphics environment},}\ }\href {\doibase 10.1109/MCSE.2007.55} {\bibfield
  {journal} {\bibinfo  {journal} {Computing in Science \& Engineering}\
  }\textbf {\bibinfo {volume} {9}},\ \bibinfo {pages} {90--95} (\bibinfo {year}
  {2007})}\BibitemShut {NoStop}%
\bibitem [{\citenamefont {Momma}\ and\ \citenamefont
  {Izumi}(2011)}]{10.1107/S0021889811038970}%
  \BibitemOpen
  \bibfield  {author} {\bibinfo {author} {\bibfnamefont {K.}~\bibnamefont
  {Momma}}\ and\ \bibinfo {author} {\bibfnamefont {F.}~\bibnamefont {Izumi}},\
  }\bibfield  {title} {\enquote {\bibinfo {title} {{{\it VESTA3} for
  three-dimensional visualization of crystal, volumetric and morphology
  data}},}\ }\href {\doibase 10.1107/S0021889811038970} {\bibfield  {journal}
  {\bibinfo  {journal} {J. Appl. Crystallogr.}\ }\textbf {\bibinfo {volume}
  {44}},\ \bibinfo {pages} {1272--1276} (\bibinfo {year} {2011})}\BibitemShut
  {NoStop}%
\bibitem [{\citenamefont {Medvedev}\ \emph {et~al.}(2017)\citenamefont
  {Medvedev}, \citenamefont {Bushmarinov}, \citenamefont {Sun}, \citenamefont
  {Perdew},\ and\ \citenamefont {Lyssenko}}]{doi:10.1126/science.aah5975}%
  \BibitemOpen
  \bibfield  {author} {\bibinfo {author} {\bibfnamefont {M.~G.}\ \bibnamefont
  {Medvedev}}, \bibinfo {author} {\bibfnamefont {I.~S.}\ \bibnamefont
  {Bushmarinov}}, \bibinfo {author} {\bibfnamefont {J.}~\bibnamefont {Sun}},
  \bibinfo {author} {\bibfnamefont {J.~P.}\ \bibnamefont {Perdew}}, \ and\
  \bibinfo {author} {\bibfnamefont {K.~A.}\ \bibnamefont {Lyssenko}},\
  }\bibfield  {title} {\enquote {\bibinfo {title} {{Density functional theory
  is straying from the path toward the exact functional}},}\ }\href {\doibase
  10.1126/science.aah5975} {\bibfield  {journal} {\bibinfo  {journal}
  {Science}\ }\textbf {\bibinfo {volume} {355}},\ \bibinfo {pages} {49--52}
  (\bibinfo {year} {2017})}\BibitemShut {NoStop}%
\end{thebibliography}%


%
%

%



\end{document}